\documentclass[twocolumn,epjc3]{svjour3}  
\journalname{Eur. Phys. J. C}

\usepackage{amsmath,amssymb,graphics,graphicx,multirow,makecell}

\newcommand{\dms}{\ensuremath{\Delta m^2}}
\newcommand{\loe}{\ensuremath{L/E}}
\newcommand{\sst}{\ensuremath{\sin^2(2\theta)}} 
\newcommand{\erec}{\ensuremath{E_{\text{rec}}}} 
\newcommand{\lrec}{\ensuremath{L_{\text{rec}}}} 
\newcommand{\loerec}{\ensuremath{\lrec/\erec}}

\newcommand{\test}     [1]{\ensuremath{\text{T}_{\text{#1}}}}
\newcommand{\hp}       [1]{\ensuremath{\text{H}_{\text{#1}}}}
\newcommand{\pvalue}   [1]{\ensuremath{\text{p}_{\text{#1}}}}
\newcommand{\PDF}   [0]{\ensuremath{h}}

\begin{document}\sloppy
\title{Statistical Methods Applied to the Search of Sterile Neutrinos}
\author{Matteo Agostini\thanksref{addr1,addr2,e1} \and Birgit Neumair\thanksref{addr1,e2}}
\institute{Physik Department E15, Technische Universit\"at M\"unchen, James-Franck-Stra\ss{}e 1, 85748, Garching, Germany \label{addr1} \and
Department of Physics and Astronomy, University College London, Gower Street, London WC1E 6BT, UK \label{addr2}
}

\thankstext{e1}{matteo.agostini@ucl.ac.uk}
\thankstext{e2}{birgit.neumair@ph.tum.de}
\date{\today}
\maketitle
\begin{abstract}
The frequentist statistical methods applied to search for short-baseline
neutrino oscillations induced by a sterile neutrino with mass at the eV scale
are reviewed and compared. 
The comparison is performed under limit setting and signal discovery scenarios,
considering both when an oscillation would enhance the neutrino
interaction rate in the detector and when it would reduce it.
The sensitivity of the experiments and the confidence regions extracted for specific
data sets change considerably according to which test statistic is used and
the assumptions on its probability distribution.
A standardized analysis approach based on the most general kind of hypothesis
test is proposed.
\end{abstract}
\section{Introduction}
A vast experimental program has been mounted in the last decade to search for a
new  elementary particle named sterile neutrino~\cite{Giunti:2019aiy}.
The sterile neutrino is a particular type of neutrino that does not interact 
through the weak force. Since B.~Pontecorvo postulated its existence in
1967~\cite{Pontecorvo:1967fh}, the sterile neutrino has become increasingly
popular and its existence is nowadays often invoked to explain the mysterious
origin of neutrino masses and dark
matter~\cite{Boyarsky:2018tvu}. The discovery of sterile
neutrinos would hence be a milestone towards the development of new theories beyond
the Standard Model, with deep repercussions in particle physics and cosmology.

The phenomenology of sterile neutrinos depends on their hypothetical mass value.
The main target of the ongoing experimental efforts is sterile neutrinos
with a mass of the order of the eV, whose existence has been hinted at
by various experiments~\cite{Aguilar:2001ty,Mention:2011rk,Giunti:2012tn} and is
still under debate~\cite{Dentler:2018sju,Giunti:2019aiy}. 
The statistical data treatment for this kind of searches
presents various challenges and has not been standardized yet. Currently,
different statistical methods are used in the field, each addressing a different
statistical question, and thus providing different results. This situation
prevents a direct comparison of the performance of the experiments and of their
outcome.  In addition, approximations often adopted in the statistical analysis
can lead to significantly inaccurate results.

In this article we review the statistical methods used in the search for
sterile neutrinos at the eV mass scale, expanding the
discussion of Refs.~\cite{Feldman:1997qc,Lyons:2014kta,Qian:2014nha} and
performing a comprehensive comparison of the analysis approaches in 
scenarios with and without a signal.
Section~\ref{sec:exp-intro} describes the phenomenology of eV-mass sterile
neutrinos, the signature sought after by the experiments, and 
the features of two toy experiments that are used in this article to compare
the analysis techniques.
Section~\ref{sec:stat-intro} reviews the statistical methods and concepts used in the
field.
The performance of the different methods are discussed in Section~\ref{sec:plr} and \ref{sec:rpl}.
Finally, in Section~\ref{sec:discussion}, the methods are compared and a standardized analysis is proposed.
 \section{Phenomenology and Experiments}
\label{sec:exp-intro}
Neutrinos of three different flavours have been observed: the electron ($\nu_e$), the muon
($\nu_{\mu}$) and the tau neutrino  ($\nu_{\tau}$)~\cite{Tanabashi:2018oca}.
These standard neutrinos can be detected by experiments because they interact
through the weak force. 
Neutrinos can change flavor as they move through space.
This phenomenon,
called neutrino flavour oscillation, is possible because neutrinos of different
flavours do not
have a fixed mass but are rather a quantum-mechanical superposition of different
mass eigenstates (i.e. $\nu_1$, $\nu_2$, and $\nu_3$), each associated to a
distinct mass eigenvalue ($m_1$, $m_2$ and $m_3$).

A sterile neutrino ($\nu_s$) would not interact through the weak force and cannot be
directly detected. However its existence would affect the
standard neutrino oscillations in two ways.
Firstly, a standard neutrino could oscillate into an undetectable sterile neutrino,
leading to a reduction of the observed event rate within the detector.
Secondly, the  mass eigenstate ($\nu_4$ with mass $m_4$) primarily associated
to the sterile neutrino would enhance the transformation probability between
standard neutrinos, leading to the detection of a neutrino flavor that is not emitted
by the source.
The experiments looking for a reduction of the interaction
rate are called ``disappearance'' experiments while the ones seeking for an
enhanced neutrino conversion are called ``appearance'' experiments.  
In principle, more than one sterile neutrino with mass at the eV scale could
exist. In this work we will focus on the scenario in which there is only one
eV-mass sterile neutrino.

\begin{figure}[tb]
   \centering
   \includegraphics[width=\columnwidth]{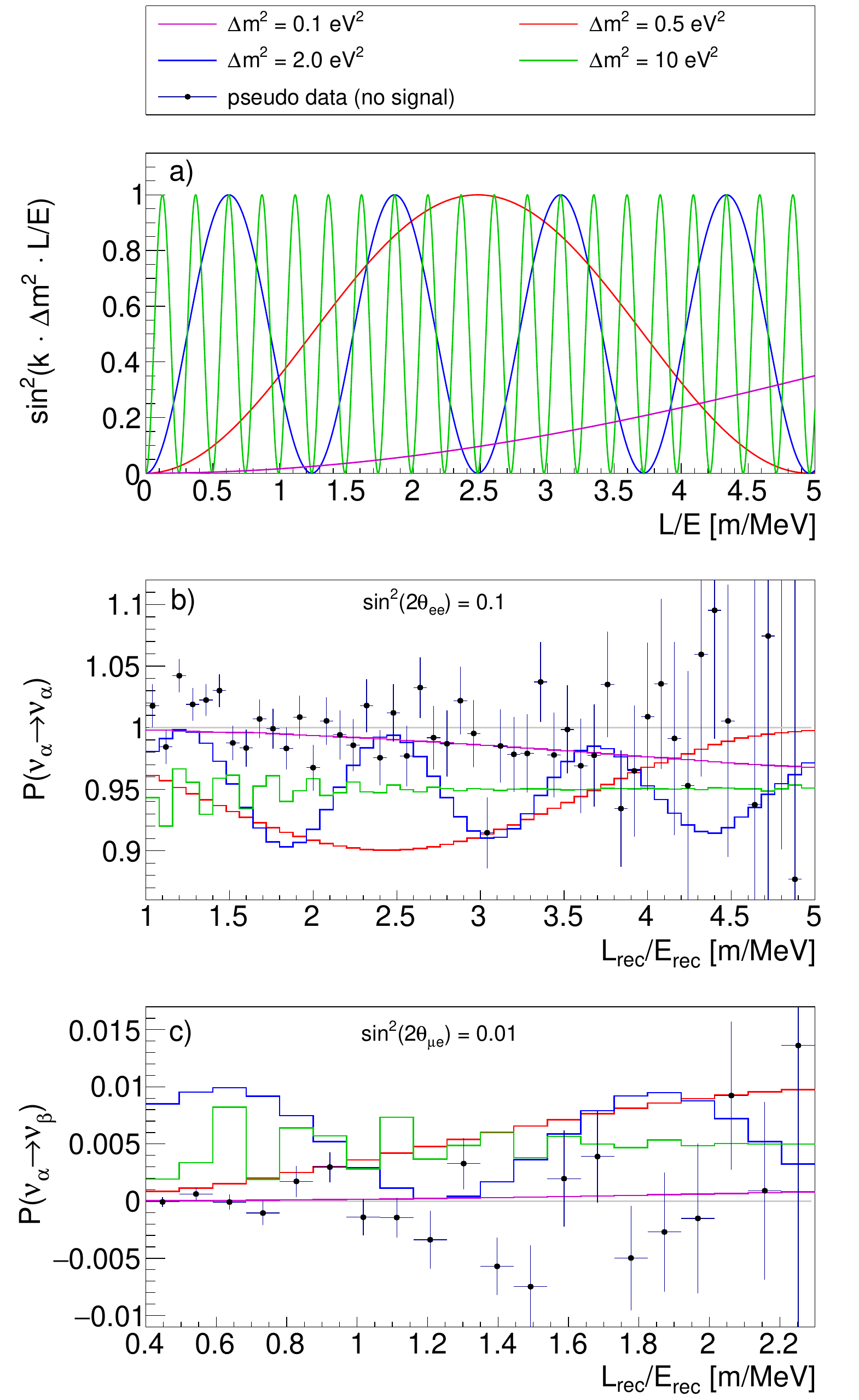}
   \caption{(a) Normalized probability of a neutrino flavor oscillation as a
   function of \loe\ for different \dms\ values. 
   The absolute probability is given by the product between the plotted normalized
   probability and \sst.
   (b and c) Probability of neutrino oscillations as a function of \loerec\
   for two toy experiments searching for a disappearance (b) and
   an appearance (c) signal.
   The probabilities are shown assuming the existence of sterile neutrinos at
   various possible \sst\ and \dms\ values.
   The reconstructed probability from pseudo-data generated with Monte Carlo
   simulations under the hypothesis that there are no sterile neutrinos are
   also shown.
   The experimental parameters of the two toy experiments are summarized in \tablename~\ref{tab:exp}. 
   The binning  reflects the typical experimental resolutions on \lrec\ and \erec.
   The error bars account for the statistical uncertainties before background subtraction.}
   \label{fig:signature}
\end{figure}

The current-generation sterile-neutrino experiments are designed to search for
oscillations among standard neutrinos at a short distance from the neutrino source,
where the effect of neutrino oscillations is expected to be negligible unless
eV-mass sterile neutrinos exist.
The oscillation probability expected by these so-called  short-baseline experiments can be
approximated by:
\begin{eqnarray}
\label{eq:osc-prob-dis}
P (\nu_{\alpha} \rightarrow \nu_{\alpha})  = &~ 1 - & \sin^2\left(2\theta_{\alpha\alpha}\right) \sin^2 \left(k \cdot \dms \cdot \frac{L}{E}\right)\\
\label{eq:osc-prob-app}
P (\nu_{\alpha} \rightarrow \nu_{\beta})  =  & & \sin^2 \left(2\theta_{\alpha\beta}\right) \sin^2\left(k  \cdot  \dms  \cdot
\frac{L}{E}\right)
\end{eqnarray}
where $P(\nu_{\alpha} \rightarrow \nu_{\alpha})$ is the survival probability for
a specific neutrino of flavor $\alpha$ and $P (\nu_{\alpha} \rightarrow
\nu_{\beta})$ is the probability for a neutrino of flavor
$\alpha$ to transform into the flavor $\beta$ ($\nu_{\alpha}$
and $\nu_{\beta}$ indicate any of the standard neutrino flavors, i.e.: $\nu_e$,
$\nu_{\mu}$ and $\nu_{\tau}$).
The mixing angles (i.e. $\theta_{\alpha\alpha}$ and $\theta_{\alpha\beta}$) and the 
difference between the squared mass eigenvalues (i.e. $\dms$) are 
physical constants\footnote{
Within the expanded oscillation phenomenology, sterile
neutrinos are described through additional non-interacting flavors,
which are connected to additional mass states via an extended
PMNS matrix. The sterile mixing angles can be expressed as a function of the
elements of the extended matrix: $\sin^2(\theta_{\alpha\beta})=4|U_{\alpha
4}|^2\left|\delta_{\alpha\beta}-|U_{\beta 4}|^2\right|$. 
The mass squared difference is typically defined as $\Delta m^2=m^2_4-m^2_1$ under the
approximation that $m_1,m_2,m_3<<m_4$. More details can be found in
Ref~\cite{Dentler:2018sju}.}.The experiments aim at extracting these constants from the measurement
of the oscillation probability as a function of the distance travelled by the
neutrino before its interaction ($L$) and its initial energy ($E$).
The maximum value of the oscillation probability is proportional to 
\sst\ that acts as a scaling factor, while the modulation
of the probability is determined by $\dms$.
The constant $k=1.27$\,MeV/eV$^2$/m 
applies when \dms\ is expressed in eV$^2$, $L$ in meters and $E$ in MeV.
The modulation of the oscillation as a function of \loe\ is shown
in \figurename~\ref{fig:signature}a  for a selection of \dms\ values.

The features of various short-baseline experiments are summarized in
\tablename~\ref{tab:exp}.
\begin{table*}[]
   \caption{Features and parameters of a selection of short-baseline experiments
      grouped according to the source of neutrinos. 
      The kind of neutrinos emitted by the source and those detected
      by the experiment are shown in the third column.
      For each experiment the accessible range of 
      \lrec, \erec, and \loerec\ are given along with the binning used to
      analyze the data and the expected number of neutrino and background events. The
      number of neutrino events is given assuming an oscillation probability of one: 
      $P (\nu_{\alpha} \rightarrow \nu_{\alpha}) =1$ in the case of disappearance experiments and 
      $P (\nu_{\alpha} \rightarrow \nu_{\beta})  =1$ in the case of appearance experiments.
      Absolute resolutions on the reconstructed baseline and energy
      ($\sigma_L$ and $\sigma_E$) are quoted
      for the mean value of \lrec\ and \erec. 
The parameters quoted in this table are sometimes effective or
      approximated quantities and are intended to give an idea of the
      signal expected in each experiment.
The last two rows show the parameters of
      two toy experiments used in this work to compare statistical methods. The
      toy experiments have parameters typical of disappearance
      searches based on nuclear reactors and appearance searches based on accelerators.}
   \label{tab:exp}
   \setlength\extrarowheight{3pt}
   \setlength{\tabcolsep}{1pt}
\begin{tabular*}{\textwidth}{lcccccccccccc}
   \hline
                                                & Detection       & Sought-after                      & \multicolumn{2}{c}{\lrec\ [m]} && \multicolumn{2}{c}{\erec\ [MeV]}  & \loerec\  [m/MeV] & binning           \\
                                                \cline{4-5} \cline{7-8} 
                                                & technology      & oscillation                       & range       & $\sigma_L$       && range       & $\sigma_E$          & range         & \lrec$\times$\erec   & $ \bar N^e$ & $\bar N^b$\\                      
      \hline
      \multicolumn{5}{l}{\it       Experiments with neutrinos from nuclear reactors:}\\
~DANSS~\cite{Alekseev:2018efk}         & Gd-coated plastic scint.& $\bar\nu_e\rightarrow\bar\nu_e$   & 11--13      & 0.2              && 1--7        & 0.6                 & 1.5-13         & 3$\times$24       & $10^6$  & $10^4$\\
~NEOS~\cite{Ko:2016owz}                & Gd-loaded liquid scint. & $\bar\nu_e\rightarrow\bar\nu_e$   & 24          & 1                && 1--7        & 0.1                 & 3.5--24        & 1$\times$60       & $10^5$  & $10^4$\\
~NEUTRINO-4~\cite{Serebrov:2018vdw}    & Gd-loaded liquid scint. & $\bar\nu_e\rightarrow\bar\nu_e$   & 6--12       & 0.2              && 1--6        & 0.3                 & 1--12          &  24$\times$9      & $10^5$  & $10^6$\\
~PROSPECT~\cite{Ashenfelter:2018iov}   & Li-loaded liquid scint. & $\bar\nu_e\rightarrow\bar\nu_e$   & 7--9        & 0.15             && 1--7        & 0.1                 & 1--9           & 6$\times$16       & $10^4$  & $10^4$\\
~SoLid~\cite{Abreu:2020bzt}            & Li-coated PVT scint.   & $\bar\nu_e\rightarrow\bar\nu_e$   & 6--9        & 0.05             && 1--7        & 0.2                 & 1--9           &                   & $10^5$  & $10^5$\\
~STEREO~\cite{Almazan:2018wln}         & Gd-loaded liquid scint. & $\bar\nu_e\rightarrow\bar\nu_e$   & 9--11       & 0.3              && 2--7        & 0.1                 & 1.3-5.5        & 6$\times$11       & $10^4$  & $10^4$\\
      \hline                                                                                             
      \multicolumn{5}{l}{\it        Experiments with neutrinos from radioactive sources:}\\
~BEST~\cite{Barinov:2019vmp}            & Ga radiochemical    & $\nu_e\rightarrow\nu_e$                     & 0.1--1      & 0.6              && 0.4--1.4    & --                  & 0.1--2.5       & & $10^4$ & $10^2$    \\
~SOX~\cite{Borexino:2013xxa,Gaffiot:2014aka} & liquid scint. & $\bar\nu_e\rightarrow\bar\nu_e$      & 4--12       & 0.15             && 2--3        & 0.1                   & 1.3--6.5      &                   & $10^4$ & $10^2$\\
      \hline
      \multicolumn{5}{l}{\it        Experiments with neutrinos from particle accelerators:}\\
~JSNS$^2$~\cite{Ajimura:2017fld}       & Gd-loaded liquid scint.   & $\bar\nu_\mu\rightarrow\bar\nu_e$ & 24        & 5                && 10--50      & 5                  & 0.5--2.5      &                    & $10^5$ & $10^2$\\
~LSND~\cite{Aguilar:2001ty}            & liquid scint.   & $\bar\nu_\mu\rightarrow\bar\nu_e$ & 30          & 10               && 20--60      & 3                   & 0.5--1.5       &1$\times$10         & $10^4$ & 30   \\
~MiniBooNE~\cite{Aguilar-Arevalo:2018gpe}      & mineral oil           & \makecell{$\nu_\mu\rightarrow\nu_e$\\$\bar\nu_\mu\rightarrow\bar\nu_e$}

                                                                  & 500         & 50               && 200--3000    & 20                   & 0.2--2.5       & 1$\times$11        & \makecell{$10^5$\\$10^4$} & \makecell{$10^3$\\$10^2$}\\   
~SBN@FNAL~\cite{Antonello:2015lea}     &  liquid-Ar TPC& $\nu_\mu\rightarrow\nu_e$
                                                                  & 110--600    & 50               && 200--3000   & 15                  & 0.1--3         & & $10^5$ & $10^3$ \\    
      \hline
      \multicolumn{5}{l}{\it        Toy experiments:}\\
~Disappearance                         &                & $\bar\nu_e\rightarrow\bar\nu_e$    & 7--10       & 0.5              && 2--7        & 0.1                   & 1--5           & 6$\times$10        & $10^5$ & $10^4$\\
~Appearance                            &                & $\nu_\mu\rightarrow\nu_e$          & 500--550    & 50               && 200--1200   & 10                  & 0.4--2.4       & 1$\times$20        & $10^5$ & $10^3$\\

   \hline
   \end{tabular*}
\end{table*}
 Different kinds of neutrino sources and detection
technologies are used. The most common neutrino sources are nuclear
reactors (producing electron anti-neutrinos up to 10\,MeV), 
radioactive sources (electron neutrinos and anti-neutrinos up to a few MeV),
and particle accelerators (muon neutrinos and anti-neutrinos up to several GeV).
The detector designs are
very different, but they mostly rely on scintillating materials and 
light detectors, or on liquid-argon time-projection chambers (LAr TPC)~\cite{Giunti:2019aiy}.

In order to extract the sterile neutrino parameters, both $L$ and $E$ must be
reconstructed for each detected neutrino.
The oscillation baseline $L$ is well
defined as either it is much larger than the dimensions of the source and the
detector -- as in accelerator-based experiments -- or the source is relatively
compact and the detector is capable of reconstructing the position of an event
-- as in experiments with radioactive isotopes or reactors. The reconstruction of
the event position is achieved with the physical segmentation of the
detector and/or with advanced analysis techniques based on the properties of the
scintillation or ionization signal.

The strategy used to reconstruct $E$ varies according to the primary 
channel through which neutrinos interact in the detector. 
Experiments using low energy anti-neutrinos can measure $E$ through a
calorimetric approach thanks to the fact that anti-neutrinos interact via 
inverse beta decay and their energy is entirely absorbed within the
detector. 
In experiments with high energy neutrinos interacting through charged-current
quasi-elastic reactions, $E$ is estimated from the kinematic 
of the particles produced in the interaction.
Some experiments measure neutrinos that interact through electron scattering and
release only a random fraction of their energy inside the detector. In this
cases the energy cannot be accurately reconstructed and monoenergetic neutrino
sources are typically used.
In the following we will use \lrec\ and \erec\ to refer to the
reconstructed value of baseline and energy. 

\tablename~\ref{tab:exp} shows for each experiment the range and resolution
of \lrec\ and \erec. To maximize the sensitivity
to sterile neutrino masses at the eV-scale, the experiments are designed to
be sensitive to \loerec\ values of the order of 1\,m/MeV.
The experiments can thus observe multiple oscillations within the detector
for \dms\ values at the eV scale. 
As the sought-after signal is similar among the experiments, the issues and challenges related to the
statistical data treatment are the same.

The analysis of an experiment can exploit two complementary pieces of
information.  When the neutrino energy spectrum, flux and cross section 
are accurately known, the integral number of neutrino
interactions expected within the detector can be computed for a given
oscillation hypothesis and compared with the observed one.
This approach is often called ``rate'' analysis.
Alternatively, the relative change of rate as a function of the interaction position and 
neutrino energy can be compared with the expectations under different
oscillation hypotheses, leaving unconstrained the integral number of events. 
This second approach is known as ``shape'' analysis. 
Rate and shape analysis are used simultaneously to maximize the
experimental sensitivity, however they are affected by different systematic
uncertainties and for a
specific experiment only one of the two might be relevant.
In the following we will discuss these two analyses separately. 
Results for a specific experiment can be estimated by interpolating between
these two extreme cases.
Experiments based on nuclear reactors sometimes use the so-called ``ratio''
method~\cite{Almazan:2018wln}, in which the energy spectrum
measured in a given
part of the detector is normalized against what observed in a reference section.
The ratio method has features similar to the shape analysis and it is not
explicitly considered in the following. 

Two toy experiments are used in this work to compare different analysis techniques.
The first one is an example of disappearance experiment representative of the
projects using nuclear reactors or radioactive isotopes as anti-neutrino source.
In these experiments, the electron anti-neutrinos partially convert into sterile
neutrinos with different probabilities as a function of \lrec\ and \erec. 
The anti-neutrino energy spectrum is considered between 2 to 7\,MeV and the
range of oscillation baselines accessible by the experiment is from 7 to 10\,m
(\loerec $=$ 1--5\,m/MeV with a resolution varying between 5 and 10\%).
The second toy experiment is an example of appearance experiment in which muon
neutrinos transform into electron neutrinos with a probability enhanced by the
existence of sterile neutrinos. 
In this case, typical for experiments based on particle accelerators,
the neutrino energy varies between 200 and 1200\,MeV
and the oscillation baseline between 500 and 550\,m 
(\loerec $=$ 0.4--2.4\,m/MeV with a resolution varying between 10 and 25\%).  

The toy disappearance experiment can observe an oscillatory pattern in the event rate as a
function of both energy and baseline, whereas the appearance experiment can
observe it only in energy as the baseline can be regarded as fixed. 
The energy distribution of both the neutrinos emitted by the source and the
background events is assumed to be flat.
This is not the case in real experiments where the initial neutrino energy
is peaked at some value and the background
spectrum depends on the background sources. However this approximation does not affect
the study reported in this work.
The experimental parameters of our toy experiments, including the number of
signal and background events, are summarized in the last two rows of
\tablename~\ref{tab:exp}. The uncertainties on the signal and background rates 
are assumed to have the realistic values of 10\% and 5\% for the appearance
experiment, and of 2\% both for the signal and background rate in the disappearance experiment.

An example of the oscillation probability reconstructed from our toy
experiments is shown in \figurename~\ref{fig:signature}b and
\ref{fig:signature}c as a function of \loerec. 
The probability is reconstructed from a set of pseudo-data generated with Monte
Carlo simulations from a model with no sterile neutrinos. 
The oscillation probability expected assuming the existence of sterile neutrinos
is shown for four mass values (\dms$=$ 0.1, 0.5, 2 and 10\,eV$^2$).
Given the \loerec\ resolution of our toy experiments,
an oscillatory pattern can be observed only for \dms\ of 0.5 and 2\,eV$^2$.
For higher \dms\ values, the frequency of the oscillation becomes too high
and only an integrated change of the rate is visible. For smaller \dms\
values, the oscillation length approaches the full \loerec\ range to which the
experiment is sensitive, resulting in a loss of sensitivity. 
In the appearance experiments the discrimination power among different oscillatory
patterns relies only on \erec\ since \lrec\ is fixed.
 
In this work we focus on short-baseline experiments. 
We do not consider other oscillation experiments (e.g. 
Daya~Bay~\cite{An:2016luf}, Double~Chooz~\cite{Hellwig:2017xhq},
RENO~\cite{Yeo:2017ied}, MINOS~\cite{Adamson:2017uda}, NOvA~\cite{Adamson:2017zcg} and
Ice~Cube~\cite{Aartsen:2017bap}) 
for which the oscillation probability
cannot be approximated by equations~\eqref{eq:osc-prob-dis} and
~\eqref{eq:osc-prob-app} as it is either complicated by the overlap between
oscillations driven by multiple mass eigenstates or by matter
effects~\cite{Dentler:2018sju}.
We also do not consider approaches that are not based on oscillations
such as the study of cosmological structures~\cite{Aghanim:2018eyx},
 the high-precision spectroscopy of beta-decays (e.g. KATRIN~\cite{Angrik:2005ep}),  or electron captures (e.g. ECHO~\cite{Gastaldo:2016kak}).
The statistical issues of
these searches are different from those of the short-baseline experiments and
would require a specific discussion.
 \section{Statistical Methods} \label{sec:stat-intro}
The goal of short-baseline experiments is to search for a signal due to
a sterile neutrino with mass at the eV-scale by measuring the oscillation
probability at different $L$ and $E$ values. The parameters of interest
associated to the sterile neutrino are the mixing angle and its mass eigenvalue.
However, because of the functional form of equations~\eqref{eq:osc-prob-dis} and
~\eqref{eq:osc-prob-app}, the observables of the experiments are a function of
the angle and mass, i.e.: \sst\ and \dms. In the following we will refer to \sst\ and \dms\ as
the parameters of interest of the analysis.

The role of statistical inference applied to the data from sterile neutrino
searches can be divided into four tasks:
\begin{enumerate}
   \item point estimation: the computation of the most plausible value for
      \sst\ and \dms;
   \item hypothesis testing: given a hypothesis on the
      value of \sst\ and \dms, decide whether to accept or reject it
      in favor of an alternative hypothesis. Among the different tests that can
      be performed, testing the hypothesis that there is no sterile neutrino
      signal (i.e.  $\sst=0$ or $\dms=0$) is of primary interest for an
      experiment aiming at a discovery;
\item interval estimation: construct a set of \sst\ and \dms\ values that
      includes the true parameter values at some predefined confidence level;
   \item goodness of fit: estimate if the data can be described by the
      model.
\end{enumerate}
The statistical methods used by sterile neutrino experiments
are based on the likelihood function.
The point estimation is carried out
using maximum likelihood estimators, i.e. by finding the values of \sst\ and
\dms\ that correspond to the maximum of the likelihood function. 
The hypothesis testing is based on the ratio of likelihoods. 
The interval estimation is carried out by inverting a set of
likelihood-ratio based hypothesis tests,
and grouping the hypotheses that are accepted.
The goodness-of-fit test can be carried out assuming the most plausible value of
the parameters of the model (i.e. the maximum likelihood estimator for \sst\ and
\dms) and using for instance a Pearson $\chi^2$ or a ``likelihood ratio''
test~\cite{Beaujean:2011zza,Baker:1983tu}. 

While the procedures for point estimation and goodness of fit are not
controversial,
the hypothesis testing differs significantly among the experiments since
multiple definitions of the hypotheses are possible.
Changing the hypothesis definition does not only affect the outcome of the hypothesis
test but also of the interval estimation, which is performed by running a set of
hypothesis tests. The comparison of tests based on different hypothesis
definitions is the subject of Sections~\ref{sec:plr}, \ref{sec:rpl} and
\ref{sec:discussion}. 

In this section we review the
ingredients needed to build the tests and the statistical concepts that will be
used in the following. Firstly, we consider the likelihood function and derive a
general form that can be applied to all experiments
(Section~\ref{sec:stat-likelihood}). Then we discuss the possible hypothesis
definitions and the resulting test statistics (Section~\ref{sec:test}).
The properties of the test statistic probability distributions are described in
Section~\ref{sec:tprob}.
Finally, in Section~\ref{sec:interval} we examine the construction of confidence regions
and in Section~\ref{sec:power} the concept of power of a test and sensitivity.

Bayesian methods have not been applied in the search for sterile neutrinos so
far. Even if their usage could be advantageous, we will not consider them 
in the following and keep the focus on the methods that are currently in use.

\subsection{The Likelihood Function}\label{sec:stat-likelihood}
Short-baseline experiments measure the oscillation baseline and the energy of
neutrinos, i.e. a pair of $\{\lrec,\erec\}$ values for each event.
\lrec\ and \erec\ are random variables whose probability distributions depend on
the true value of $L$ and $E$. Monte Carlo simulations are used to construct the probability
distributions of \lrec\ and \erec\ for a neutrino event given a \sst\ and \dms\ value, 
$p_{e}(L,E|\sst,\dms)$, and for a background event, $p_{b}(L,E)$.
Additional quantities are sometimes measured, however they are ultimately used
to constrain the background or the systematic uncertainties and can be
neglected in this work.

To our knowledge, all the experiments organize the data in histograms and base
their analysis on a binned likelihood function. The use of histograms is
motivated by the fact that the number of neutrino
events is large, between $10^4$ and $10^6$ as shown in
\tablename~\ref{tab:exp}.
Binning the data leads to a new set of random variables that are the numbers of observed
events in each bin: $N^{obs} = \{N^{obs}_{11}, N^{obs}_{12}, \cdots,
N^{obs}_{ij}\cdots\}$ where $i$ runs over the $L_{rec}$ bins and $j$ over the
$E_{rec}$ bins.
Consistently, we indicate with $\PDF^e_{ij}$  and $\PDF^b_{ij}$ the integral
of the probability distribution function for neutrino and background events over
each bin: 
\begin{eqnarray}
   \PDF^{e}_{ij} &=& \int_{L,E\in bin_{ij}} p_{e}(L,E|\sst,\dms)\,dL\,dE\\
   \nonumber \\
   \PDF^{b}_{ij} &=& \int_{L,E\in bin_{ij}} p_{b}(L,E)\,dL\,dE.
\end{eqnarray}
The generic likelihood function can hence be written as: 
\begin{eqnarray}
   {\mathcal{L}}(\sst, \dms, N^e, N^b|N^{obs}) = 
   \prod_{ij} \mathcal{P}(N^{obs}_{ij} | 
\nonumber \\
   N^e \cdot  
   \PDF^e_{ij}(\sst, \dms) + 
   N^b \cdot   \PDF^b_{ij})
\end{eqnarray}
where $i$ and $j$ run over \lrec\  and  \erec\ bins,
$\mathcal{P}(N|\lambda)$ indicates the Poisson probability of measuring $N$
events given an expectation $\lambda$, and $N^e$ and $N^b$ are scaling factors
representing the total number of standard neutrino and background events.

External constraints on the number of neutrino and background events related to
auxiliary data are in this work
included as additional multiplicative Gaussian terms:
\begin{eqnarray}
   \mathcal{{L}} \rightarrow  \mathcal{{L}} 
   \cdot\,\mathcal{G}(\bar N^e(\sst,\dms)|N^e ,\sigma^e) 
   \cdot\,\mathcal{G}(\bar N^b           |N^b ,\sigma^b)  \nonumber\\
   \label{eq:likelihood-tot}
\end{eqnarray}
where $\mathcal{G} (\bar N|N,\sigma)$ indicates the probability of measuring
$\bar N$ given a normal distributed variable with mean $N$ and standard
deviation $\sigma$. The pull terms can be based on other probability distributions
(e.g. log-normal or truncated normal distributions), however their specific
functional form is not relevant for our study.
It should be noted that the expected number of neutrino counts
$\bar N^e$ depends on the particular oscillation hypothesis tested.  Examples of the likelihood can be found in \ref{sec:app-likelihood}.

While \sst\ and \dms\ are the parameters of interest of the analysis, $N^e$ and
$N^b$ are nuisance parameters. 
The constraints on these parameters could follow different probability
distributions and additional nuisance parameters could also be needed
to account for systematic uncertainties in the detector response, neutrino
source, and event reconstruction efficiency.
The actual number of nuisance parameters and the particular form of their
constraints in the likelihood does not affect the results of our work. 
Systematic uncertainties typically cannot mimic the expected oscillatory signal,
even though they can change the integral rate.
Thus, in a pure rate analysis a precise understanding of the systematic
uncertainties including those related to the background modeling is mandatory.    
 
To keep the discussion general, in the following
we will indicate with $\boldsymbol{\eta} = \{N^e, N^b,...\}$
a generic vector of nuisance parameters. Each parameter has an
allowed parameter space, for instance the number of neutrino and background
events
are bounded to non-negative values. The nuisance parameters are assumed
to be constrained in their allowed parameter space even if not explicitly
stated.

The general form of the likelihood given in equation~\eqref{eq:likelihood-tot}
accounts for a simultaneous rate and shape analysis. 
A pure shape analysis will be emulated by removing 
the pull term on the number of neutrino events.
Conversely, a pure rate analysis will be emulated by enlarging the size of the bins in
$\PDF^e$ and $\PDF^b$ up to the point at which there is a single bin and any
information on the number of events as a function of \lrec\ or \erec\ is lost.

\subsection{Hypothesis Testing and Test Statistics}\label{sec:test}
The hypothesis testing used nowadays in particle physics is based on the approach
proposed by Neyman and Pearson in which the reference 
hypothesis $\hp{0}$ (i.e. the null hypothesis) is compared against an
alternative hypothesis $\hp{1}$~\cite{casella2003statistical}. 
The test is a procedure that specifies for which data sets the decision is made to accept
\hp{0} or, alternatively, to reject \hp{0} and accept \hp{1}.
Usually a hypothesis test is specified in terms of a test statistic \test{}
and a critical region for it. The test statistic is a function of the data that
returns a real number. The critical region is the range of test statistic
values for which the null hypothesis is rejected.

The critical region is chosen prior the analysis such that the test rejects \hp{0} when \hp{0} is actually true
with a desired probability. This probability is
denoted with $\alpha$ and called the ``size'' of the test. 
In the physics community, it is more common to quote $1-\alpha$ and refer to
it as the ``confidence level'' of the test. 
For instance, if \hp{0} is rejected with $\alpha=5\%$ probability when it is true, 
the test is said to have 95\% confidence level (C.L.). 
In order to compute the critical thresholds, the probability distribution of the test
statistic must be known.   In our work, the distributions are constructed from
large ensembles of pseudo-data sets generated via Monte Carlo techniques. 

In sterile neutrino searches a hypothesis is defined by a set of allowed
values for \sst\ and \dms. The null hypothesis is defined as:
\begin{eqnarray}
   \hp{0}:\{\sst, \dms:\sst=X, \dms =Y\}
\end{eqnarray}
where $X$ and $Y$ are two particular values. 
Since the mixing angle and the mass eigenvalue are defined as non-negative
numbers by the theory and $m_4\geq m_1$, the most general version of the
alternative hypothesis is 
\begin{eqnarray}
   \hp{1}:\{\sst, \dms: 0\leq\sst\leq 1, \dms \geq 0\}.
   \label{eq:thetaT2}
\end{eqnarray}
A test based on these two hypotheses leads to a generalized likelihood-ratio test
statistic of the form~\cite{casella2003statistical}:
\begin{equation}
\test{} = -2\ln \dfrac
{\sup\limits_{          \boldsymbol{\eta}} \mathcal{L}\left(\sst=X,\dms=Y,\boldsymbol{\eta} | N^{obs}\right)}
{\sup\limits_{\sst,\dms,\boldsymbol{\eta}} \mathcal{L}\left(\sst  ,\dms ,\boldsymbol{\eta} | N^{obs}\right)}
\label{eq:t2}
\end{equation}
where the denominator is the maximum of the likelihood for the observed data set over the
parameter space allowed for the parameters of interest ($\{\sst, \dms\}\in \hp{1}$)
and the nuisance parameters.
The numerator is instead the maximum of the likelihood in the restricted space in
which \sst\ and \dms\ are equal to the value specified by \hp{0}.

If the value of \dms\ or \sst\ are considered to be known because of theoretical
predictions or of a measurement, then the parameter space of the alternative
hypothesis can be restricted.
Restricting the parameter space is conceptually equivalent to folding into the analysis
new assumptions and changes the question addressed by the hypothesis test. 
The smaller is the parameter space allowed by the alternative hypothesis, the
greater the power of the test will be.

Three tests have been used in the context of sterile neutrino
searches and are summarized in \tablename~\ref{tab:stat}.
\begin{table*}[]
   \caption{Definition of the test statistics used for sterile-neutrino searches
      in the presence of nuisance parameters ($\boldsymbol{\eta}$). 
      The null hypothesis is $\hp{0}:\{\sst, \dms: \sst=X, \dms =Y\}$ for all
      tests while the alternative hypothesis \hp{1} changes. The free parameters
      of interest in \hp{1} are shown in the second column.  The name of the
      techniques based on each test statistics and a selection of experiments
      using them are listed in the last columns.}
      \label{tab:stat}
      \centering
     \begin{tabular}{cc c c c c}
         \hline
        \makecell{Test Statistic Computed for \\ $ \hp{0}:\{\sst, \dms: \sst=X,
        \dms =Y\}$}
        & \makecell{Free Parameters\\ of Interest}  
        & \makecell{Associated \\ Names} 
        & Experiments \\  
         \hline \\
         $ \test{2} = -2\ln \dfrac
         {\sup\limits_{          \boldsymbol{\eta}} \mathcal{L}\left(\sst=X,\dms=Y,\boldsymbol{\eta} | N^{obs}\right)}
         {\sup\limits_{\sst,\dms,\boldsymbol{\eta}} \mathcal{L}\left(\sst  ,\dms ,\boldsymbol{\eta} | N^{obs}\right)} $
         & \sst , \dms
         & \makecell{2D Scan or \\ global p-value}
         & \makecell{LSND, MiniBooNE,\\ PROSPECT}\\[30pt]

         $ \test{1} = -2\ln \dfrac
         {\sup\limits_{          \boldsymbol{\eta}} \mathcal{L}\left(\sst=X,\dms=Y,\boldsymbol{\eta} | N^{obs}\right)}
         {\sup\limits_{\sst,\boldsymbol{\eta}} \mathcal{L}\left(\sst  ,\dms=Y ,\boldsymbol{\eta} | N^{obs}\right)} $
         & \sst
         & \makecell{Raster Scan or\\ local p-value}
         & \makecell{NEOS, STEREO}\\[30pt]

         $\test{0} = -2\ln \dfrac
         {\sup\limits_{\boldsymbol{\eta}} \mathcal{L}\left(\sst=X,\dms=Y,\boldsymbol{\eta} | N^{obs}\right)}
         {\sup\limits_{\boldsymbol{\eta}} \mathcal{L}\left(\sst=0  ,\dms=0 ,\boldsymbol{\eta} | N^{obs}\right)} $
         & ---
         & \makecell{Simple Hypthosis Test \\ or Gaussian CL$_\text{s}$}
         & \makecell{DANSS}\\
         \hline

      \end{tabular}
\end{table*}
 The most general test is the one that we just described and that leads
to the test statistic given in equation~\eqref{eq:t2}. We will indicate this
test statistic with \test{2}.
This test is agnostic regarding the value of \sst\ or \dms\ and can be applied to search
for a sterile neutrino with unknown parameters.

The second test statistic used in the field can be traced back to the situation in which 
the mass squared difference is considered to be perfectly known and is equal to the
value of the null hypothesis ($\dms=Y$). In
this case the alternative hypothesis and its related test statistic are:
\begin{equation}
\hp{1}:\{\sst, \dms: 0\leq\sst\leq 1, \dms=Y\}
\end{equation}\begin{equation}
\test{1} = -2\ln \dfrac
{\sup\limits_{          \boldsymbol{\eta}} \mathcal{L}\left(\sst=X,\dms=Y,\boldsymbol{\eta} | N^{obs}\right)}
{\sup\limits_{\sst,\boldsymbol{\eta}} \mathcal{L}\left(\sst  ,\dms=Y ,\boldsymbol{\eta} | N^{obs}\right)}.
\label{eq:t1}
\end{equation}While the numerator of \test{1} is the same of \test{2},
the maximum of the likelihood at the denominator is now computed over a narrower
parameter space, restricted by the condition $\dms=Y$.

The third test corresponds to the simplest kind of hypothesis test that can be
performed. Both the null and alternative hypothesis have the parameters of
interest fully defined.
The alternative hypothesis is now the no-signal hypothesis
and this leads to a test of the form:
\begin{equation}
\hp{1}:\{\sst, \dms: \sst=0, \dms=0\}
\end{equation}
\begin{equation}
\test{0} = -2\ln \dfrac
{\sup\limits_{\boldsymbol{\eta}} \mathcal{L}\left(\sst=X,\dms=Y,\boldsymbol{\eta} | N^{obs}\right)}
{\sup\limits_{\boldsymbol{\eta}} \mathcal{L}\left(\sst=0  ,\dms=0 ,\boldsymbol{\eta} | N^{obs}\right)}.
\label{eq:t0}
\end{equation}
The numerator and denominator are  the maximum likelihoods for fixed
values of the parameters of interest, where the maximum is computed over the
parameter space allowed for the nuisance parameters. 
By construction, the no-signal hypothesis is always accepted when it is used as
\hp{0}, since the test statistic becomes identically equal to zero.

Nowadays, the value of \sst\ or \dms\ is still considered to be unknown and 
all the parameter space accessible by the experiment is probed in search for a signal.
This situation should naturally lead to the usage of \test{2}.
However the maximization of the likelihood required by \test{2} is challenging
from the computational point of view. Reducing the dimensionality of the
parameter space over which the likelihood is maximized can enormously simplify
the analysis and, for such a practical reason, \test{1} and \test{0}
are used even if the restriction of the parameter space is not intended.

In the neutrino community, the analysis based on \test{2} has been called  ``2D
scan'' or ``global scan'' while the analysis based on \test{1} is known as ``raster
scan''~\cite{Feldman:1997qc,Lyons:2014kta}. 
In the absence of nuisance parameters, the definition of these test statistics
reduce to those discussed in Ref.~\cite{Feldman:1997qc}.
\test{0} has been used in the framework of a method called ``Gaussian
CLs''~\cite{Qian:2014nha}.

The search for new particles at accelerators presents many similarities
with the search for sterile neutrinos. For instance, in the search for the Higgs
boson, the sought-after signal is a peak over some
background. The two parameters of interest are the mass of the Higgs boson, which
defines the position of the peak, and its coupling with other particles, which
defines the amplitude. 
Similarly, in the search for sterile neutrinos \dms\ defines the shape
of the signal and \sst\ its strength. 
When the Higgs boson is searched without assumptions on its mass and coupling,
a test similar to \test{2} is performed (i.e. a ``global p-value'' analysis).
When the mass is assumed to be known, a test similar to \test{1} is used (i.e. a
``local p-value'' analysis)~\cite{ATLAS:2011tau}. 
Procedures for converting
a local into a global p-value analysis have been developed in the last
years~\cite{Gross:2010qma,Ranucci:2012ed} and are nowadays used to avoid the direct usage of
\test{2} that is computationally demanding. This procedure is 
known as a correction for the ``look-elsewhere effect''.
We have studied the correction described in Ref.~\cite{Gross:2010qma} 
and found that it
does not provide accurate results for sterile neutrino experiments
because of the oscillatory nature of the
sought-after signature. Our studies are discussed in 
\ref{sec:app-lee}.

\subsection{Test Statistic Probability Distributions}\label{sec:tprob}
The test statistic \test{2} and \test{1} can assume any non-negative value.
If the absolute maximum of the likelihood corresponds to \hp{0}, these test
statistics are identically zero. The farther the absolute maximum is from the
parameter space of the null hypothesis, the larger the test statistic value
becomes. If the null hypothesis is true, the probability distribution of this
kind of test statistic is expected to converge to a chi-square function in the
large sample limit, but only if the regularity conditions required by Wilks'
theorem are met~\cite{Wilks:1938dza}.
In particular, given the ratio between the dimensionality of the parameter space
for the null and alternative hypothesis (i.e. the number of free parameters of
interest in the likelihood maximization), \test{2} would converge to a
chi-square with two degrees of freedom
and \test{1} to a chi-square with one degree of freedom. 
As discussed in  Section~\ref{sec:plr-wilks}, the conditions required by Wilks'
theorem are not always valid in sterile neutrino experiments and the assumption that
the test statistic follows a chi-square distribution can lead to significantly
inaccurate results.

The probability distributions of \test{0} are qualitatively different from those
of \test{2} and \test{1}. 
\test{0} is negative when the tested signal hypothesis is more likely than the no-signal hypothesis, 
positive in the opposite case. 
The larger is the test statistic value, the more the tested
hypothesis is disfavoured. 
Under mild conditions, the probability distribution of \test{0} converges
to a Gaussian function~\cite{Qian:2014nha}.

Our results are based on test statistic probability
distributions constructed from ensembles of pseudo-data.
Firstly a grid in the \sst\ vs. \dms\ space is defined. Secondly, for each point on
the grid an ensemble of pseudo-data is generated. The probability distributions
are hence constructed by computing the test statistic for the pseudo-data
in the ensemble. The pseudo-data are generated for a fixed value of the nuisance
parameters.
More details on our procedure are described in \ref{sec:app-pseudo-data}. 

\subsection{Interval Estimation and Confidence Regions}\label{sec:interval}
The results of a neutrino oscillation search are generally summarized by
a two-dimensional confidence region in the \sst\ vs. \dms\ space. The
confidence region defines a set of parameter values that are compatible with
the data given a certain confidence level. The construction of a
confidence region is formally referred to as an interval estimation.

One of the most popular statistical techniques to construct a confidence region is
through the inversion of a set of hypothesis tests~\cite{casella2003statistical,stuart2009kendall}.
This is also the technique used by experiments searching
for sterile neutrinos. The construction starts from the selection of a specific
test and its resulting test statistic. The parameter space considered for 
\hp{1} is naturally the space in which the region will be defined. 
Usually a grid is fixed over this space and a test is run for each point. The
tests in this set have the same \hp{1} but \hp{0} is changed to the value of the parameters
at each point.
This standard construction guarantees that the properties of
the test statistic carry over to the confidence region and the confidence level
of the region is equal to that of the test~\cite{casella2003statistical}. 

Since  the confidence region is constructed in the parameter space considered by
the alternative hypothesis, tests based on \test{2}  would naturally lead to
two-dimensional confidence regions in the \sst\ vs. \dms\ space, tests based on
\test{1} to one-dimensional regions in the \sst\ space, and tests based on \test{0} to
point-like regions. As we already mentioned, 
\test{0} and \test{1} are used even if the restriction of the parameter space is
not intended. To build two-dimensional regions for \test{0} and \test{1} a
non-standard procedure is used, i.e. the final region is created 
as union of one-dimensional or point-like regions.
For instance, while \test{1} would require the value of \dms\ to be known, one can
technically construct a
one-dimensional \sst\ region for a scan of ``known'' \dms\ values and then
take the union of these regions. Assuming that the true \dms\ is among the
scanned values, the confidence level of the union will be the same of the test.

The standard procedure for constructing confidence regions ensures that
inverting uniformly most powerful tests provides uniformly most accurate
confidence regions, i.e. regions with minimal probability of false
coverage~\cite{casella2003statistical}. 
This is not true for the non-standard procedure described above, which 
indeed produces regions with peculiar features and pathologies as discussed in
Sections~\ref{sec:plr} and \ref{sec:rpl}.

\subsection{Power of the Test and Sensitivity}\label{sec:power}
The performance of the different kinds of hypothesis tests can
be studied by comparing their expected outcome under the assumption that an
hypothesis is true.
The idea of expected outcome is captured by the statistical concept of ``power'' of
a test. The power is defined as the probability that the test rejects the null
hypothesis when the alternative hypothesis is true~\cite{casella2003statistical}. 

In high-energy physics the concept of power is replaced by the 
idea of sensitivity of an experiment. The sensitivity is defined as the set of
hypotheses for which the test would have a 50\% power. The focus is thus shifted
from the  expected outcome of a test to a set of hypotheses.
Two kinds of sensitivities are commonly used. The exclusion sensitivity 
is the set of hypotheses that have a 50\% chance to be excluded assuming there is no
signal (\hp{0} is the oscillation signal for which the test has a 50\% power).
The discovery sensitivity is the set of hypotheses that, if true, 
would allow in 50\% of the experiments to reject the no-signal hypothesis (\hp{0} is  now the no-signal
hypothesis).  
More details on how the sensitivities are defined and computed can be found in \ref{sec:app-neyman}.

Both sensitivities will be displayed as contours in the \sst\ vs. \dms\ parameter
space and for a 95\% C.L. test. A larger confidence level is typically required
for a discovery, however we prefer to use the same value for the exclusion and
discovery sensitivity to ease their comparison.
 \section{2D and Raster Scan} \label{sec:plr}
In this section we compare the confidence regions built using \test{2} and
\test{1}. 
The comparison is done using the toy experiments introduced in
Section~\ref{sec:exp-intro}: a disappearance experiment representative of
searches based on reactor neutrinos and an
appearance experiment representative of the accelerator-based experiments.
First we focus on the sensitivity of the toy experiments
(Section~\ref{sec:plr-sensitivity}) and then consider
the results extracted for specific sets of data (Section~\ref{sec:plr-brazilianflag}). 
Finally, in Section~\ref{sec:plr-wilks}, we study the impact of 
approximating the test statistic distributions with chi-square functions.
Our results and conclusions fully agree with previous
works~\cite{Feldman:1997qc,Lyons:2014kta}.

\subsection{Sensitivity}
\label{sec:plr-sensitivity}
\begin{figure*}[]
   \centering
   \includegraphics[width=1\textwidth]{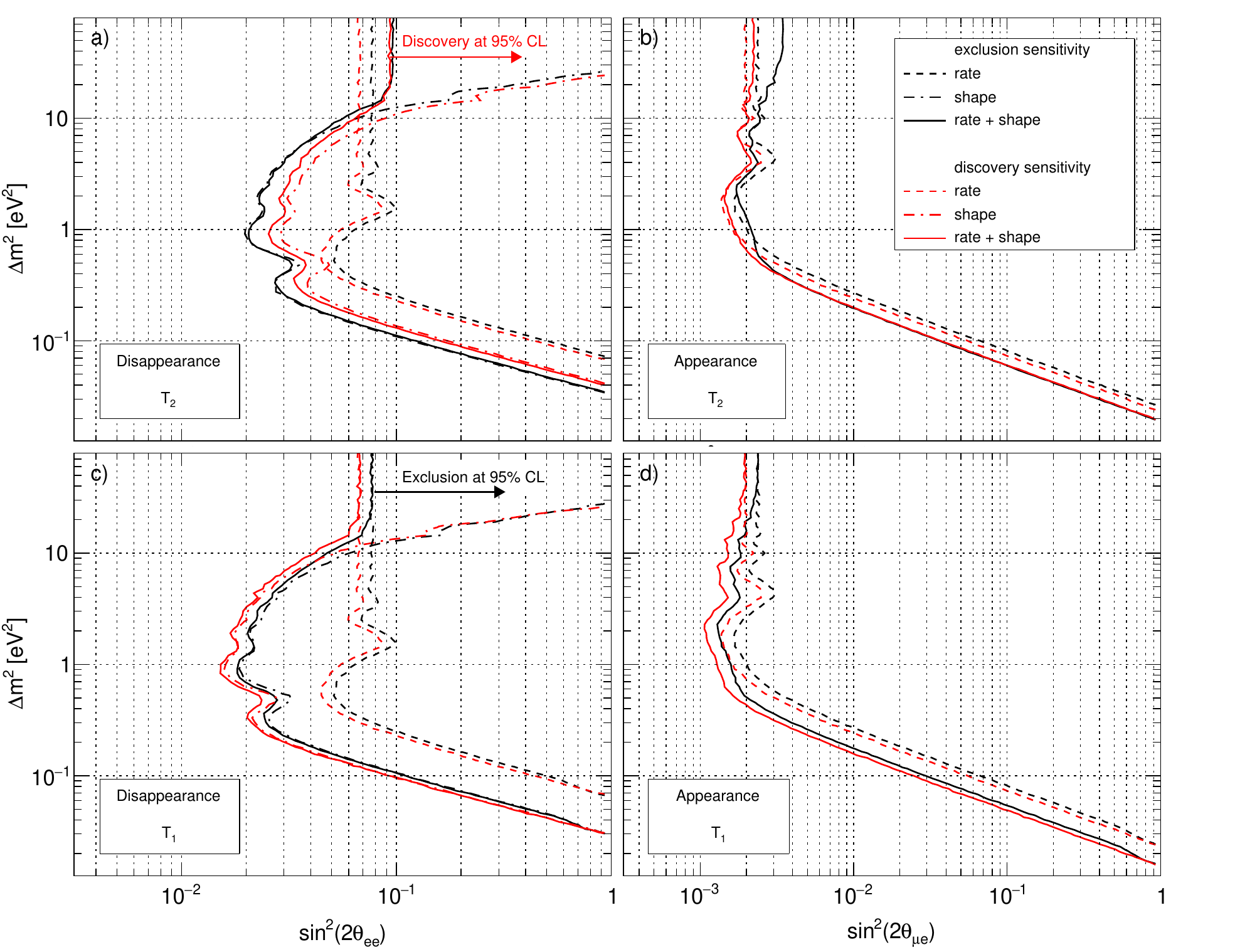}
   \caption{Exclusion and discovery sensitivity at 95\% C.L. 
      for a toy disappearance and appearance experiment. 
      The sensitivities are shown for the test statistics \test{2} and \test{1}. 
      Whenever possible, the contribution of the shape and rate analysis are
      displayed separately. The discovery sensitivity is typically associated to
      a larger confidence level compared to the exclusion sensitivity, however
      here we use the same value to highlight their differences.}
   \label{fig:sensitivity}
\end{figure*}

The exclusion and discovery sensitivities of our toy
disappearance experiment based on the statistic \test{2} are shown in
\figurename~\ref{fig:sensitivity}a. 
The exclusion sensitivity (black lines) delimits the parameter space that has
a 50\% chance to be rejected by a 95\%-C.L. test under the
assumption that sterile neutrinos do not exist.
The discovery sensitivity (red lines) delimits the set of
hypotheses which, assuming those to be true, have a 50\% chance 
that the no-signal hypothesis is rejected by a 95\%-C.L. test.
The figure shows separately the sensitivity for a rate and shape analysis
(dotted lines) that are useful to illustrate which kind of information contributes 
most to the overall sensitivity as a function of \dms. Three \dms\ regions can be
identified in \figurename~\ref{fig:sensitivity}a:
\begin{itemize}
\item $\dms>~10$ eV$^2$: the oscillation length is smaller than the detector
   resolution on \lrec\ and/or \erec, making the experiment sensitive only to an
   overall reduction of the integral rate (sensitivity dominated by the rate
   analysis);
\item 0.1\,eV$^2 < \dms <10$\,eV$^2$: the oscillation length is larger than the
   experimental resolution and smaller than the range of \lrec\ and/or \erec\
   values accessible by the detector, making the experimental sensitivity
   dominated by the shape analysis;
\item $\dms <0.1$\,eV$^2$: the oscillation length becomes larger than
   the detector dimensions. The experimental sensitivity decreases with the \dms\
   value (i.e. with increasing oscillation length) and larger \sst\ values are
   needed in order to observe a signal. The sensitivity is approximately
   proportional to the product $\sst\times\dms$.
\end{itemize}
Example of the expected oscillations in these three regions are shown in
\figurename~\ref{fig:signature}b and \ref{fig:signature}c.

The total sensitivity is given by a non-trivial combination of the
sensitivity of the rate and shape analysis.
The rate and shape analysis are emulated by considering only
parts of the likelihood function (see Section~\ref{sec:stat-likelihood}) that
would otherwise have common parameters.
The sensitivity for the rate analysis is higher than the
total one for high \dms\ values.
This feature is related to the fact that \sst\ and \dms\ are fully degenerate
parameters in a rate analysis based on \test{2}. A rate analysis uses a single
piece of information and cannot fix two correlated parameters.  
Given an observed number of events, the global maximum of the
likelihood function can be obtained for infinite combinations of \sst\ and
\dms\ values.  The degeneracy is however broken when the rate information is
combined with the shape one. The number of effective degrees of freedom of
the problem changes, and this results in a reduction of sensitivity.

The sensitivities of our toy appearance experiment computed for \test{2} is shown
in \figurename~\ref{fig:sensitivity}b.
The same \dms\ regions discussed for the disappearance experiment can be
identified, even if the relative weight of the shape and rate information is
different.
In particular, in the appearance searches a shape analysis can provide information on \dms\ but not on \sst. The number of expected $\nu_e$ events is indeed proportional to the product of the oscillation amplitude \sst\ and flux of $\nu_\mu$ neutrinos. If the flux is left
unconstrained in the fit, no statement can be made about the oscillation amplitude. 
This is the reason why the shape analysis contribution is not displayed.
The rate analysis accounts for the bulk of the sensitivity and adding the 
shape information does not result in a net improvement, in the sense that the
slight improvement is compensated by the reduction of sensitivity due to the
increased number of effective degrees of freedom discussed above.
Having a sensitivity dominated by the rate analysis is typical for experiments
using accelerators as neutrino source. 

The exclusion and discovery sensitivities are similar to each
other for both the disappearance and appearance experiments. Some differences
are however present. When computing the discovery sensitivity, the 
hypothesis tested is the no-signal hypothesis. Since $\sst=0$ or $\dms=0$ are
points at the edge of the allowed parameter space, the number of degrees
of freedom of the problem decreases when testing them and the power of the test increases. 
The exclusion sensitivity is instead computed for values of the
parameters far from the edges.
This is the reason why the discovery sensitivity is in general
expected to be stronger than that the exclusion one.
However the situation is reversed in the shape analysis because of a peculiar
feature of the sterile neutrino signature.  Statistical fluctuations between
bins mimic an oscillation signal and the maximum of the likelihood is always
found far from the no-signal hypothesis. 
This decreases significantly the power of the test for a discovery while it
does not affect much the exclusion case. If the shape analysis
dominates the overall sensitivity, as in our disappearance experiment, 
its features propagate also to the combined sensitivities. 

\figurename~\ref{fig:sensitivity}c and \ref{fig:sensitivity}d show the
sensitivities for our toy disappearance and appearance experiments computed for \test{1}.
The overall features are similar to those of \test{2} and the weight of the rate
and shape information in the three \dms\ regions are also consistent.
However, since the parameter space of the alternative hypothesis is now
restricted, \test{1} has greater power than \test{2} for a given \dms\ value.
This leads to  sensitivities that are stronger by up to a factor 2 in terms of \sst.
This is particular evident for high \dms\ values where, differently from
\test{2}, now the number of effective degrees of freedom in the alternative
hypothesis is always one (only \sst\ is free) and the total sensitivity is
equal to the one of the rate analysis.
The restriction of the parameter space is also the reason why the maximum of the
likelihood can now correspond to the no-signal hypothesis and the discovery sensitivity
is stronger than the exclusion sensitivity even in a shape
analysis.  

It should be emphasized that the mixing angle in the plot for 
the disappearance experiment is different from that of the appearance experiment. 
The minimal value of \sst\ accessible by an experiment cannot
be used as a figure of merit to compare disappearance and appearance experiments.
A comparison can however be done assuming a specific theoretical model that
connects the value of  $\sin^2(2\theta_{ee})$ and $\sin^2(2\theta_{\mu
e})$~\cite{Dentler:2018sju}. 

\subsection{Results from Observed Data Sets}\label{sec:plr-brazilianflag}
The confidence region derived from an observed data set can significantly differ from
the expectations because of statistical fluctuations on the number of
signal and background events. This issue is particularly relevant when no signal
is observed and an upper limit on a parameter is reported. Frequentists upper
limits can indeed become extremely strong in case of background fluctuations.

In sterile neutrino searches, when no signal is observed, 
the confidence region extends down to $\sst=0$ for most of the \dms\ values and
it is bounded by an upper limit on \sst\ that plays the role of the maximum signal strength. 
It is hence informative to report the observed upper limit along with its
expected value and variance under the no-signal hypothesis. This has been
first proposed in Ref.~\cite{Feldman:1997qc} and it is nowadays common practice.

\begin{figure}[]
   \centering
   \includegraphics[width=\columnwidth]{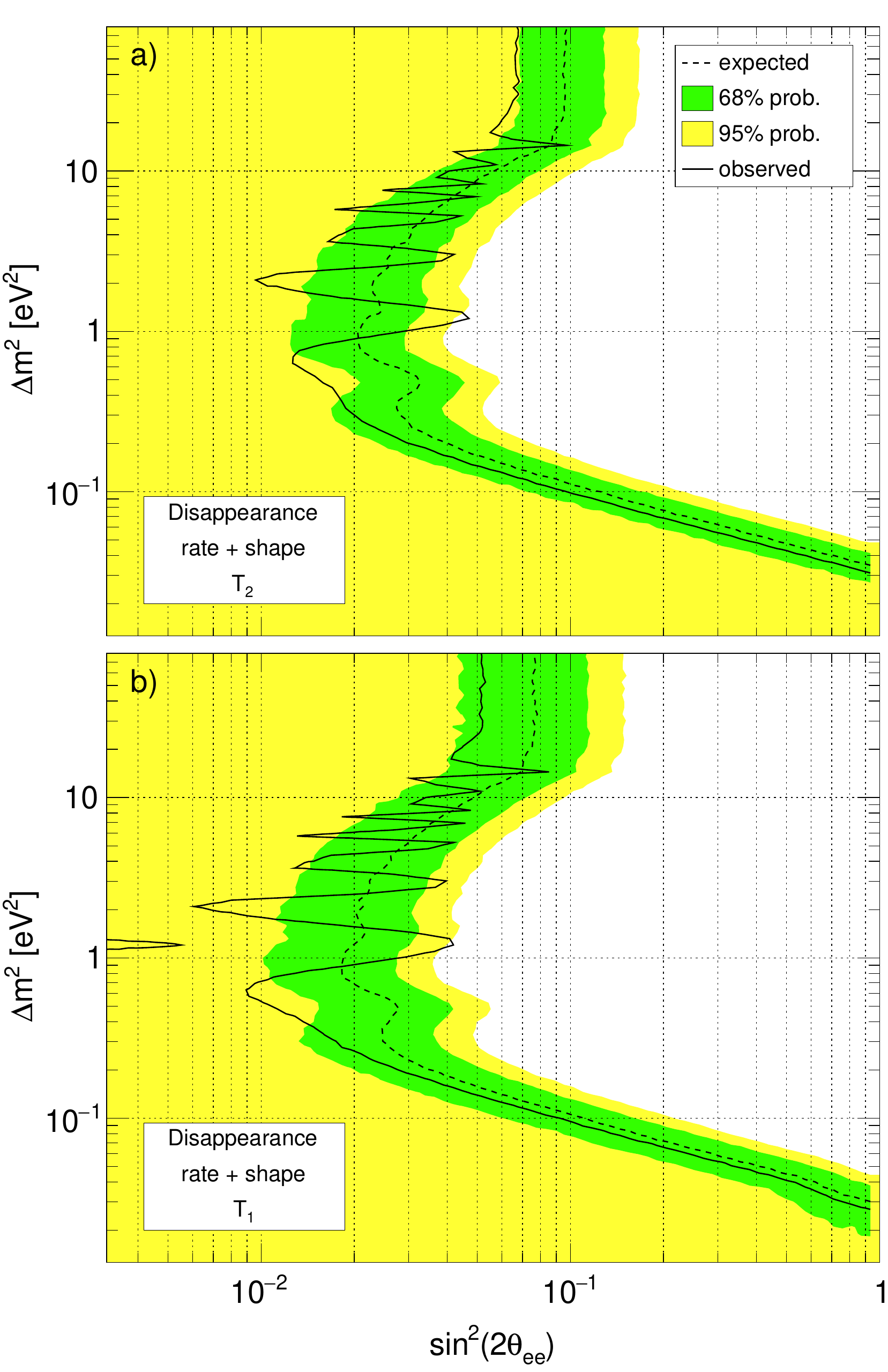}
   \caption{Confidence regions at 95\%-C.L. for a
      pseudo-data set generated by the toy disappearance experiment under the
      no-signal hypothesis. The top plot
      is obtained using \test{2} while the bottom plot using \test{1}.
      The probability distribution of the upper bound of the confidence region
      expected under the no-signal hypothesis is displayed through its median
      value (i.e. the exclusion sensitivity) and the 68\% and 95\%
      central intervals.}
   \label{fig:brazflag}
\end{figure}

\figurename~\ref{fig:brazflag}a shows the confidence region derived with
\test{2} from a pseudo-data set generated for the toy disappearance
experiment under the no-signal hypothesis. In addition to the confidence region,
the expected distribution of the upper limit is displayed in terms of its median
value and 68\%/95\% central intervals.  
The median is exactly the exclusion sensitivity plotted in
\figurename~\ref{fig:sensitivity}a.
The observed upper limit fluctuates around the median expectation.
This is true for all possible realizations of the data as
the likelihood is maximized for a specific phase of the oscillatory pattern that matches
the statistical fluctuations between the bins of the data set.
This phase is reproduced at regularly spaced values of \dms\ over the full
parameter space.
The limit gets weaker when the phase helps describing the data, stronger
when it does not. The overall shift of the observed limit with respect to the
median value is instead due to the fact that the random number of events injected in this particular 
data set is slightly above its median expectation. The width of the green and
yellow bands gives an idea of the magnitude of the fluctuations at a given \dms, as
they contain the upper limit on \sst\ with a probability of 68\% and 95\% respectively. 

The results and expectations based on \test{1} are shown in
\figurename~\ref{fig:brazflag}b.
For a given \dms\ value, \test{1} has greater power than \test{2} as
the parameter space allowed under the alternative hypothesis is smaller.
This leads to stronger limits in terms of \sst.
On the other hand, the non-standard construction of the confidence region
can lead to accept the no-signal hypothesis (i.e. $\sst=0$) at some \dms\
value and reject it at others (see for instance \dms$\sim$1\,eV$^2$).
This can happen because the tests performed by \test{1}
at different \dms\ values are independent by each other.

\begin{figure}[]
   \centering
   \includegraphics[width=\columnwidth]{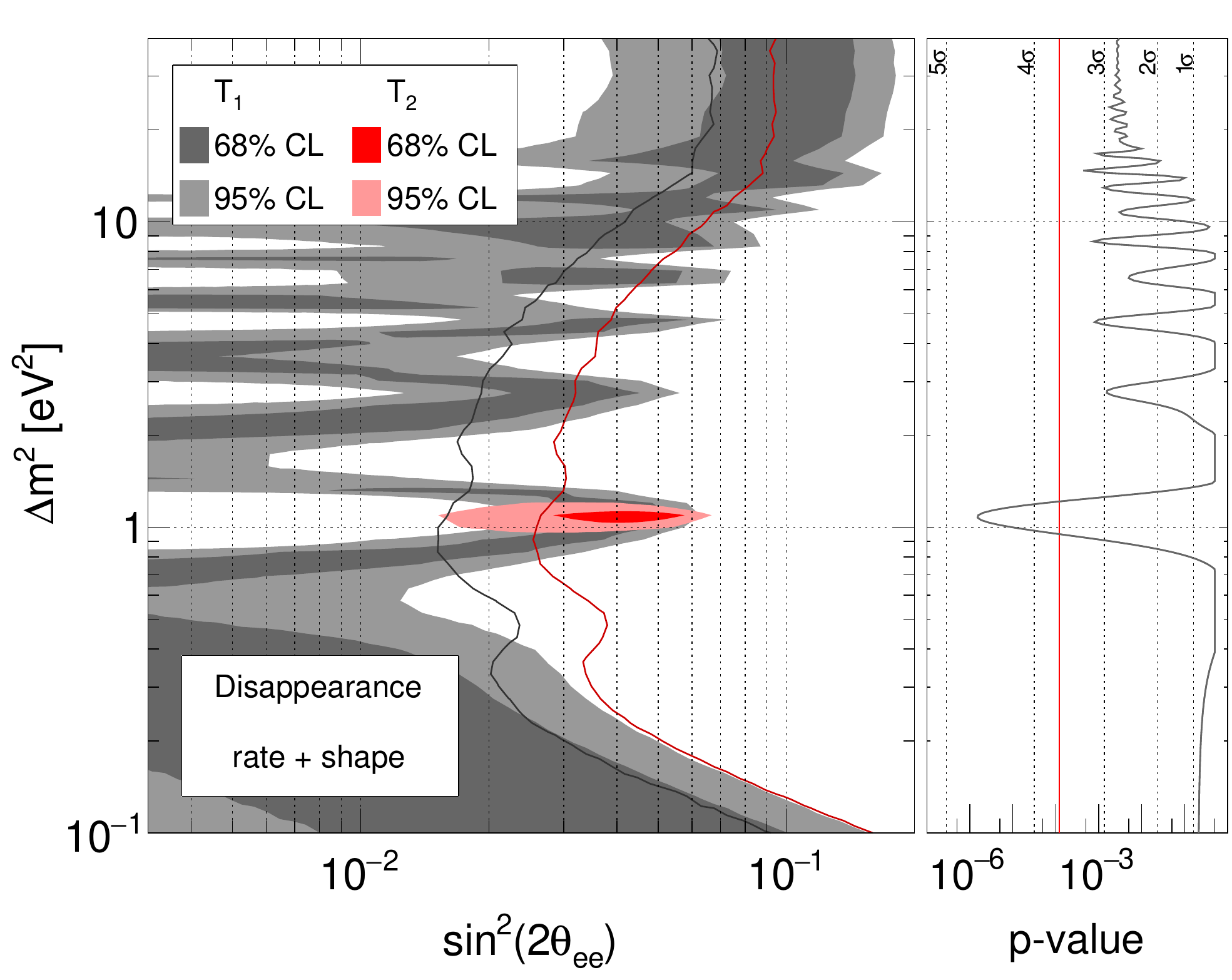}
   \caption{Confidence regions at 68\% and 95\% C.L. for pseudo-data generated
      by the toy disappearance experiment assuming the
      existence of a sterile neutrino with $\sst=0.04$ and $\dms=1$\,eV$^2$. 
      The 95\% C.L. discovery sensitivity is shown
      for both the statistic \test{2} (red line) and \test{1} (grey line).
      The right panel shows the minimal size of a test based on \test{1} that is
      required to reject the no-signal hypothesis as a function of the \dms\
      value (grey line). The minimal test size for \test{2} (red line) is
      independent by \dms\ as all values are tested simultaneously. Differently
      from \test{2}, \test{1} would reject the no-signal hypothesis even for a test size of
      $10^{-5}$ (corresponding to a 4$\sigma$ two-sided Gaussian deviation).
      This discrepancy is due to the look-elsewhere effect. }
   \label{fig:discovery}
\end{figure}

The difference between \test{2} and \test{1} is more evident when a signal is present in the data. 
\figurename~\ref{fig:discovery} shows the reconstructed confidence regions
for a pseudo-data set generated assuming a sterile
neutrino with $\sst=0.04$ and $\dms=1$\,eV$^2$. The confidence
regions are shown for 68\% and 95\% C.L. along with the discovery sensitivity.
The analysis based on \test{2} is able to properly pin down the signal and 
it returns a two-dimensional confidence region surrounding the true parameter values.
\test{1} returns a \sst\ region that is similar to that of \test{2}
for \dms\ values close to the true one.  However it returns
an allowed region for any \dms\ value. 
This is again due to the non-standard construction of the confidence region used in
combination with \test{1}. As tests performed at different \dms\ values are
independent from each other, an allowed \sst\ interval is always found.

In summary, the greater power of \test{1} in terms of \sst\ comes at the cost
of losing any capability in constraining \dms. This is consistent with the fact
that this test statistic originates from an hypothesis test in
which \dms\ is considered to be a known and fixed parameter.

\subsection{Validity of Wilks' theorem}\label{sec:plr-wilks}
Constructing the probability distributions of the test statistic through Monte
Carlo methods can be computationally challenging and often it is avoided by
invoking the asymptotic properties of the generalized likelihood ratio test.
If the regularity conditions required by Wilks' theorem are
met~\cite{Wilks:1938dza,Algeri:2019arh}, \test{2} and \test{1} are indeed expected to follow a chi-square
distribution with a number of degrees of freedom equal to the effective number
of free parameters of interest~\cite{Cowan:2010js}.
Sterile neutrino experiments do not typically fulfill these regularity
conditions.
The parameter \sst\ is often reconstructed next to the border of its allowed range (only positive values
are physically allowed) or far from its true value when the statistical
fluctuations mimic a signal. This induces a bias in its maximum likelihood
estimator. In addition, for \test{2}, the alternative hypothesis becomes
independent by \dms\ when \sst\ is equal to zero.

The impact of assuming Wilks' asymptotic formulas has been evaluated by studying the coverage probability, i.e. the probability that the confidence region covers the true
value of the parameters of interest~\cite{casella2003statistical}.
If the asymptotic formulas are a good approximation of the actual test statistic
distribution, the coverage should be  equal to the confidence level of the test
used to create the confidence region. A direct comparison 
of the test statistic distributions is discussed in \ref{sec:app-pdf}.

\begin{figure*}[]
   \centering
   \includegraphics[width=\textwidth]{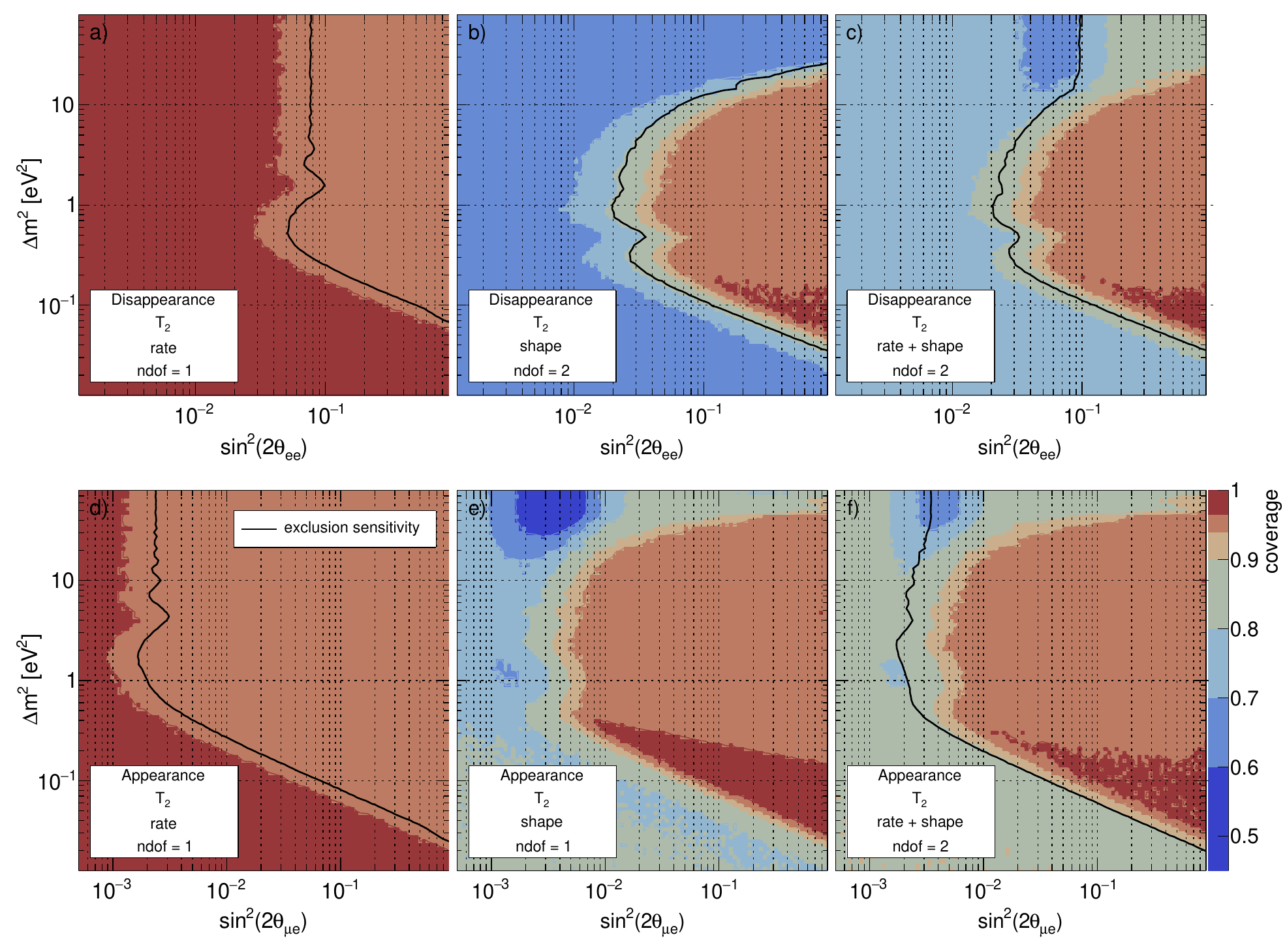}
   \caption{Coverage of the 95\%-C.L. confidence region based on the test performed
    with \test{2} and assuming that this test statistic follows chi-square distributions.
    The number of degrees of freedom of the chi-square distribution is two (i.e.
    the number of free parameters in the likelihood) unless these parameters are
    degenerate such as \sst\ and \dms\ in the rate analysis or \sst\ and $N^{e}$
    in the shape analysis of the appearance experiment.
    The top panels show the coverage probability for the toy disappearance
    experiment, the bottom panels for the toy appearance experiment. For each
    kind of experiment the coverage for a rate analysis (left), a shape analysis
    (middle) and the full analysis (right panel) are shown separately.
    The color palette used to show the coverage probability is the same in all
    plots.
    The 95\% C.L. exclusion sensitivity is also displayed to delimit
 the parameter space in which the experiment is sensitive.}
   \label{fig:coverage}
\end{figure*}

The coverage probability computed for \test{2} assuming the validity of Wilks'
theorem is shown in \figurename~\ref{fig:coverage} for both our toy
disappearance and appearance experiments, considering separately a rate and
shape analysis.
The test statistic distributions have been approximated by a
chi-square with one or two degrees of freedom, according to the number
of non-degenerate parameters of interest in the alternative hypothesis (see
insets in the figure).
The coverage is generally correct in the parameter space where the experiment is
sensitive to a signal.
The rate analysis shows just a slight overcoverage where the
experiment is not sensitive. This is expected as \sst\ is
bounded to positive values, causing an effective reduction of the degrees
of freedom of the test when the signal is reconstructed close
to the edge of the allowed parameter space~\cite{Chernoff:1954eli}.

The shape analysis has instead a severe undercoverage for \sst\ values
below the sensitivity of the experiment and the coverage can be as low as 60\%,
while its nominal value should be 95\%.
The undercoverage is connected to the fact that when a binned analysis is
performed, it is always possible to find a sterile neutrino hypothesis whose
oscillatory pattern helps reproducing the statistical fluctuations between
bins. As a result, even if no signal is present in the data, the maximum of the
likelihood always corresponds to some oscillation hypothesis. This is
conceptually equivalent to overfitting and it artificially increases the degrees of freedom of the test and the test
statistic values (see discussion in \ref{sec:app-pdf}).
A region of overcoverage is present also in the parameter space within the
sensitivity of the experiment at low \dms\ values, where  the oscillation length becomes
close to the dimension of the detector, creating a degeneracy between the
parameters of interest. Together with the restriction of the parameter space ($\sst\leq1$), the number of effective degrees of
freedom changes.

When the analysis includes both the rate and shape information, the coverage
shows a combination of the features discussed above. In particular, 
in the parameter space beyond the sensitivity of the experiment, the overcoverage
of the rate analysis partially compensates for the undercoverage of the shape
analysis. Severe undercoverage regions are however still present, consistently
with the results obtained in Ref.~\cite{Feldman:1997qc}.

\begin{figure*}[]
   \centering
   \includegraphics[width=1\textwidth]{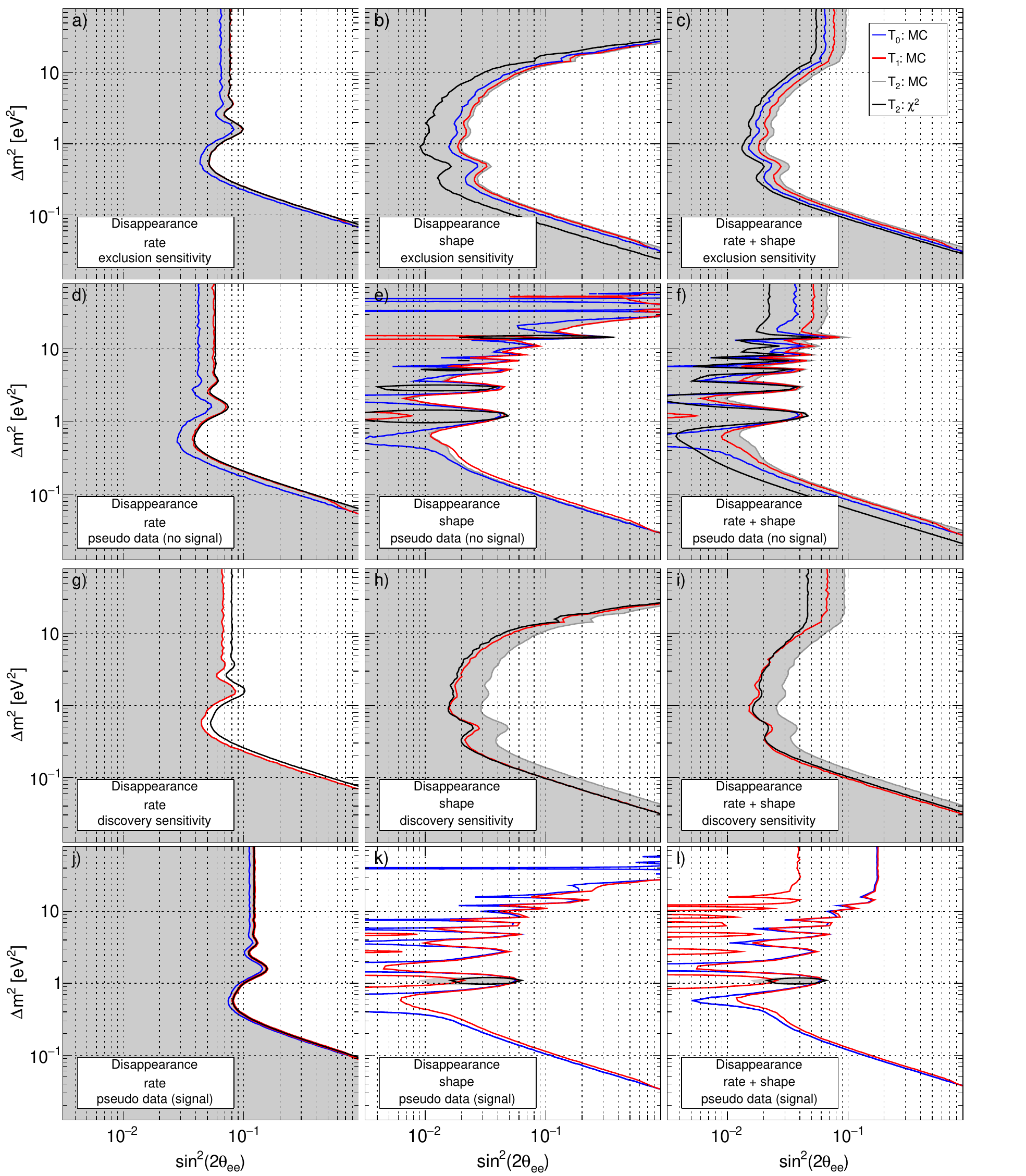}
   \caption{Comparison of the sensitivity and confidence regions at 95\% C.L. for the toy
      disappearance experiment obtained with  \test{2}, \test{1} and \test{0}.
      The exclusion and discovery sensitivities (first and third row
      respectively) have been computed using 10000 pseudo-data sets. The
      confidence regions for a concrete data-set have instead been calculated
      for a pseudo-data set generated under the no-signal hypothesis (second row) and one 
      generated assuming a sterile neutrino signal with \sst$=$0.04 and
      \dms$=$1\,eV$^2$ (fourth row).
      The rate (left) and shape analysis (middle) are shown independently and
      combined together (right column).
The probability distribution of the test statistics are computed
      using Monte Carlo techniques. Results obtained when approximating the
      distributions with chi-square functions are shown for \test{2}.
}
      \label{fig:comparison}
\end{figure*}

The difference between the outcome of a test based on probability
distributions constructed with Monte Carlo techniques and their chi-square
approximation is shown in \figurename~\ref{fig:comparison}.
Both the sensitivities and the confidence regions
reconstructed from pseudo data are significantly different, up to 70\% in terms of \sst.
For experiments with a sensitivity dominated by the shape analysis, the
confidence region can even switch from an upper limit to an island, leading to an
unjustified claim for a discovery. The probability for this event to occur can be
significant, up to 40\% in our toy experiment for the considered hypothesis.
More details on the probability distributions of \test{2} and the option to
compute exclusion sensitivities based on the Asimov data set are discussed in
\ref{sec:app-pdf}.

While the asymptotic approximation is not satisfactory for tests based
on \test{2}, it is instead very good for tests based on \test{1}.
The coverage of \test{1} has exactly the same features of
\figurename~\ref{fig:coverage}a and \ref{fig:coverage}d and therefore it is
not shown here. The coverage is correct in the region in
which the experiment is sensitive and is slightly higher (97.5\%) in the
parameter space beyond the experimental sensitivity. The possibility of avoiding
a Monte Carlo construction of the probability distributions of \test{1}
is a significant advantage and contributed to make \test{1} popular in the
sterile neutrino community.
 \section{Testing of Simple Hypotheses}\label{sec:rpl}
The exclusion sensitivity based on \test{0} for our toy disappearance experiment 
is shown in the first row of \figurename~\ref{fig:comparison} separately for the 
rate, shape and combined analysis.
This test provides a sensitivity significantly stronger than what obtained with 
\test{2} and \test{1}. This is expected as the test involves ``simple''
hypotheses in which the parameters of interest are fixed.
The parameter space of the alternative hypothesis is now even more restricted
than for \test{1} and the test has maximum power.
The discovery sensitivity cannot be calculated as the no-signal hypothesis is
always accepted when used as \hp{0}.

The confidence regions extracted for specific pseudo-data sets not containing
a signal is shown in the second row of \figurename~\ref{fig:comparison}.
\test{0} can provide extremely stringent constraints on \sst\ that are orders of
magnitudes beyond the sensitivity.  
To mitigate this behaviour this test statistic is used in combination
with the CL$_\text{s}$ method~\cite{Read:2000ru} that penalizes constraints
stronger than the sensitivity by introducing an overcoverage in the test. 
The combination of \test{0} and the CL$_\text{s}$
method is known as ``Gaussian CL$_\text{s}$''~\cite{Qian:2014nha}.

The plots in the fourth row of \figurename~\ref{fig:comparison} show the
confidence regions extracted for a pseudo-data set with an injected signal.
These regions have two peculiarities.  Similarly to \test{1}, the non-standard
construction of the confidence region produces an allowed \sst\
interval for each \dms\ value. 
However, differently from \test{1}, the \sst\ intervals are now always connected to the
no-signal hypothesis \sst$=$0, even for the true \dms\ value, as the alternative hypothesis 
in the test is now fixed to the no-signal hypothesis.

In conclusion, while \test{0} has a greater power and can produce the strongest
limits in terms of \sst, it produces confidence regions that cannot constrain
either of the parameters of interest.  We confirm that the probability
distribution of \test{0} converges to a normal distribution for our toy
appearance and disappearance experiment as reported in
Ref.~\cite{Qian:2014nha}.

 \section{Comparison and Discussion}\label{sec:discussion}
The main difference among the statistical methods applied to the search for sterile
neutrinos has been traced back to the definition of the alternative hypothesis
in the hypothesis testing procedure. The considered definitions
lead to three different test statistics that are used to construct confidence
regions in the \sst\ vs. \dms\ parameter space.
The sensitivities and confidence regions constructed for each test are compared in
\figurename~\ref{fig:comparison}.

In \test{2}, the parameter space of the alternative hypothesis covers all possible
values of \sst\ and \dms. 
This test is the natural choice when the values of the parameters
of interest are unknown and a generic search over the full parameter space is
intended.
Using this test for an interval estimation procedure provides 
naturally two-dimensional confidence regions in the \sst\ vs. \dms\ space.
The probability distributions of this test statistic are not well approximated
by chi-square functions in the analysis of sterile neutrino experiments, and
such an approximation can lead to very inaccurate confidence regions and even to
erroneously reject the no-signal hypothesis.

In \test{1}, the value of \dms\ is assumed to be known prior to the experiment
and the parameter space of the alternative hypothesis is restricted to a unique
\dms\ value.
\test{1} naturally generates one-dimensional confidence regions in the \sst\ space. 
Two-dimensional confidence regions can be technically created as as union of
\sst\ intervals, each computed for a different fixed value of \dms.
Such confidence regions have proper coverage but also some pathologies. In
particular, while the constraints on \sst\ are more stringent than for
\test{2}, the test has no capability to constrain \dms\ and the confidence region 
extends over any \dms\ value.
The conditions of Wilks' theorem are almost fulfilled and its probability distribution follows
accurately a chi-square function except in the parameter space close to the
physical border where the probability distribution becomes half a chi-square function
and half a delta-Dirac function at zero~\cite{Chernoff:1954eli}. 

The test statistic \test{0} compares two simple hypotheses with a
fixed value of the parameters of interest. The alternative hypothesis is 
defined as the no-signal hypothesis. Thus, the no-signal hypothesis is 
accepted by construction when used as \hp{0}.
The natural confidence regions constructed using this test are point-like.
Two-dimensional regions in the \sst\ vs \dms\ can be obtained as union of 
point-like confidence regions, but this non-standard construction produces regions that
do not constrain the parameters of interest and only set upper limits on \sst. 
These limits are consistently stronger than for \test{1} and \test{2} as this
test has maximum power.
The asymptotic formulas, namely Gaussian distributions, seem to describe well the probability distribution of
this test statistic in a large set of conditions.

In summary, all the test statistics are conceptually correct and have a
natural scope of application. The more information is available, the more we can
restrict the parameter space of the alternative hypotheses and the greater the
power of the test becomes. However, some issues arise when these
tests are used -- regardless of what is their natural scope -- to build two-dimensional confidence regions in the \dms\ vs. \sst\ space.
The regions produced by \test{1} and \test{0} do not constrain \dms\ as they
have an allowed \sst\ interval for any \dms\ value. 
In addition, the regions produced by \test{0} do not even constrain \sst\, as
the \sst\ interval is always connected to \sst$=$0, even in the presence of a
strong signal.
Since the primary goal of the current experiments is to find a signal at unknown
\sst\ and \dms\ values, we find natural to adopt an
analysis that is able to pin down simultaneously both oscillation parameters and
recommend the usage of \test{2}.

To ease the comparison of the performance and results from different
experiments, it would be convenient for the field to adopt a standardized
analysis procedure.
Based on the results presented in this article and expanding the proposal of
Ref.~\cite{Feldman:1997qc}, such a standard analysis could follow these steps:
\begin{enumerate}
   \item identification of the most likely value for \sst\ and \dms\ defined as
      the value corresponding to the maximum of the likelihood function over the
      space $\{\sst, \dms: 0\leq\sst\leq1, \dms\geq0\}$ (i.e. maximum likelihood
      estimators);
   \item check that the data is compatible with the model corresponding to the most likely value of the
      parameters of interest by using a
      ``likelihood ratio'' goodness-of-fit test~\cite{Baker:1983tu}
      whose probability distribution is verified or constructed with Monte Carlo
      techniques\footnote{We studied the probability distribution of the likelihood-ratio
         goodness-of-fit test for a large number of configurations of our
         disappearance experiment and found that generally it can be approximated
         with a chi-square function. Nevertheless, we also identified some
         situations, e.g. in a shape analysis, in which the distribution
         follows a chi-square function with a number of degrees of freedom different
         from the expected one.};
   \item construct the two-dimensional confidence region based on \test{2}.
      If the no-signal hypothesis is accepted, the confidence region will extend
      down to vanishing \sst\ values and its upper limit can be plotted along
      with its median value (i.e. the exclusion sensitivity) and 68/95\% central intervals expected under the
      no-signal hypothesis (as in \figurename~\ref{fig:brazflag}).  If the
      no-signal hypothesis is instead rejected, the confidence region can be
      plotted for different confidence levels along with the discovery sensitivity
      as in \figurename~\ref{fig:discovery}. 
\end{enumerate}

When the number of events observed by an experiment is large and each bin of the
data set contains tens of counts or more, the Poisson probability in the
likelihood function can be approximated by a Gaussian probability. In this case, the
likelihood function can  be converted into a chi-square function. This treatment can be regarded as a
sub-case of what is discussed in the previous sections.
However, independently by the number of events,
Wilks' theorem is not valid for \test{2} because of the presence
of physical borders and of the statistical fluctuations in the data sample that mimic 
a sterile neutrino signature. 
In addition, the alternative hypothesis becomes
independent by \dms\ when \sst\ tends to zero.
The construction of the test statistic probability distribution through Monte
Carlo techniques is hence mandatory in order to ensure accurate results.
The Monte Carlo construction is computationally
demanding, but it is feasible as proved by the experiments that are already
performing it. Indeed, the proposed analysis based on \test{2} is similar to the
one used by e.g. MiniBooNE and PROSPECT.

The inapplicability of Wilks' theorem and the non-trivial interplay between the rate and shape analysis 
have repercussions also on the global fits for which
the likelihood of each experiment must be combined and
the probability distribution of \test{2} must be computed by generating simultaneously
pseudo-data for all the experiments considered.
For this reason, it would be useful if the experiments
would release in addition to their likelihood fit function and their data, also the
probability distributions of the individual signal and background components used in the fit and for
the pseudo-data generation.
 \section{Conclusions}
The statistical methods used to search for short-baseline neutrino oscillations
induced by an hypothetical sterile neutrino with mass at the eV scale have been
reviewed and compared. 
Three hypothesis testing procedures are used in
the field to create confidence intervals. Each procedure is based on a specific
test statistic.  We identified how two out of the three tests make implicit assumptions
on the value of the parameters of interest for sterile neutrinos, i.e. \dms\
and \sst. Making different assumptions changes the question addressed by the
test and, consequently, changes the result of the analysis.

For the first time, the performance of the three tests
have been compared in a coherent way over a comprehensive set of limit setting and
signal discovery scenarios. In particular, we considered both disappearance and appearance
experiments as well as rate- and shape-based analyses. For each scenario
we constructed the probability distributions of the test statistic using Monte
Carlo techniques and found that they can differ significantly from the usual
asymptotic approximations. The confidence regions reconstructed by the three
tests can be significantly different, making hard to compare the results from
experiments that adopt different analyses.
Our results are consistent with those obtained
in Refs.~\cite{Feldman:1997qc,Lyons:2014kta,Qian:2014nha}
for specific scenarios.

The current generation sterile-neutrino
searches aim at finding a signal anywhere in the parameter space available to the experiment. 
This should naturally lead to an analysis
based on a test without any assumption on the value of the oscillation
parameters. In this way, the analysis will be able to constrain both the value
of \dms\ and \sst\ when a signal is observed. 
Thus, we recommend the use of \test{2} and the
construction of its probability distributions through Monte Carlo techniques.
 \appendix
\section{The Likelihood}\label{sec:app-likelihood}
The general form of the likelihood used in the analysis of sterile neutrino
experiments is given in Section~\ref{sec:stat-likelihood}. 
The computational task of finding the maximum of the likelihood is typically performed by
minimizing the negative logarithm  of the likelihood (NLL). 
Moving to the logarithm space is
convenient from the computational and numerical point of view. This is one of
the reason why the test statistics are defined as the logarithm of the likelihood.
\begin{figure}[tb]
   \centering
   \includegraphics[width=\columnwidth]{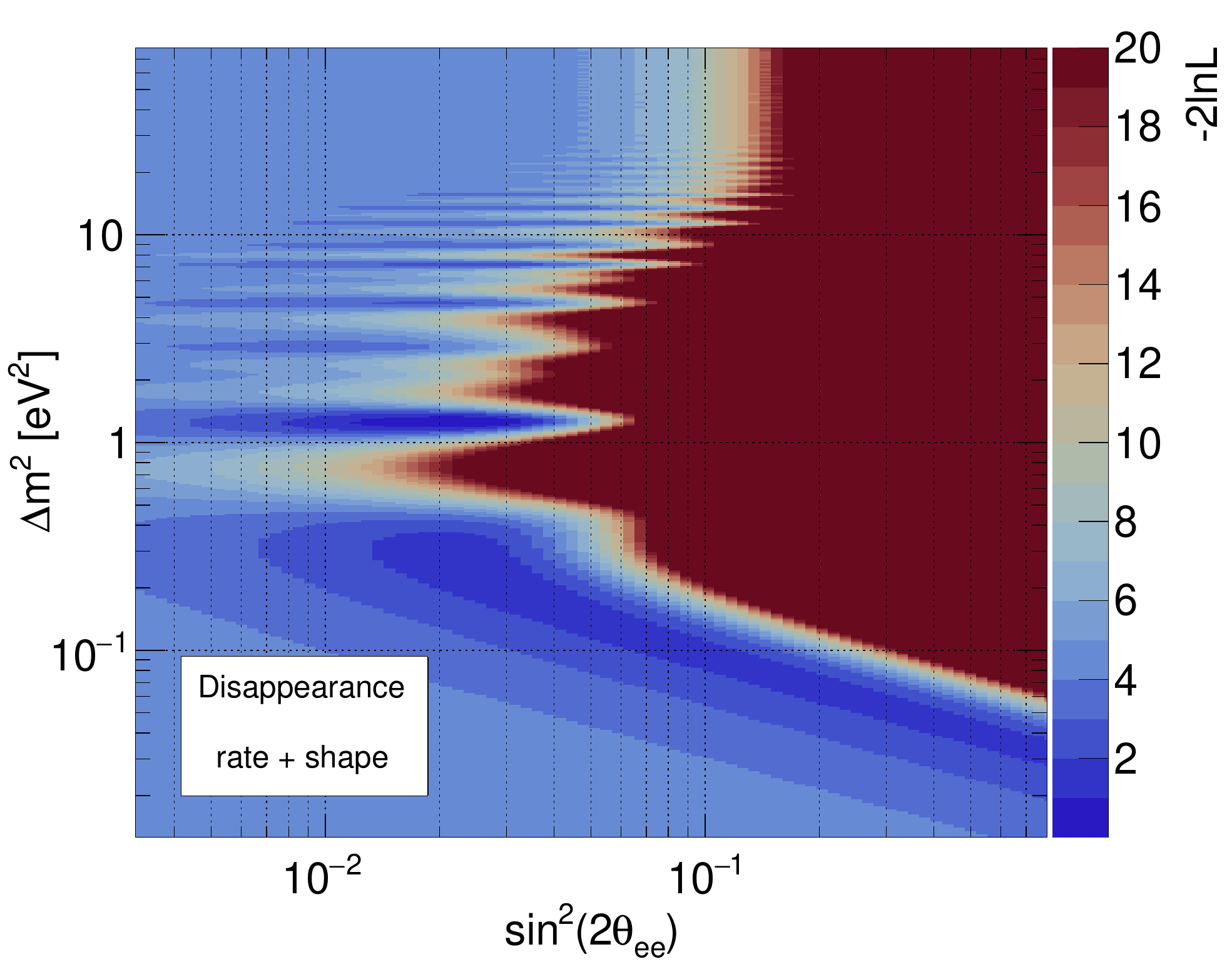}
   \includegraphics[width=\columnwidth]{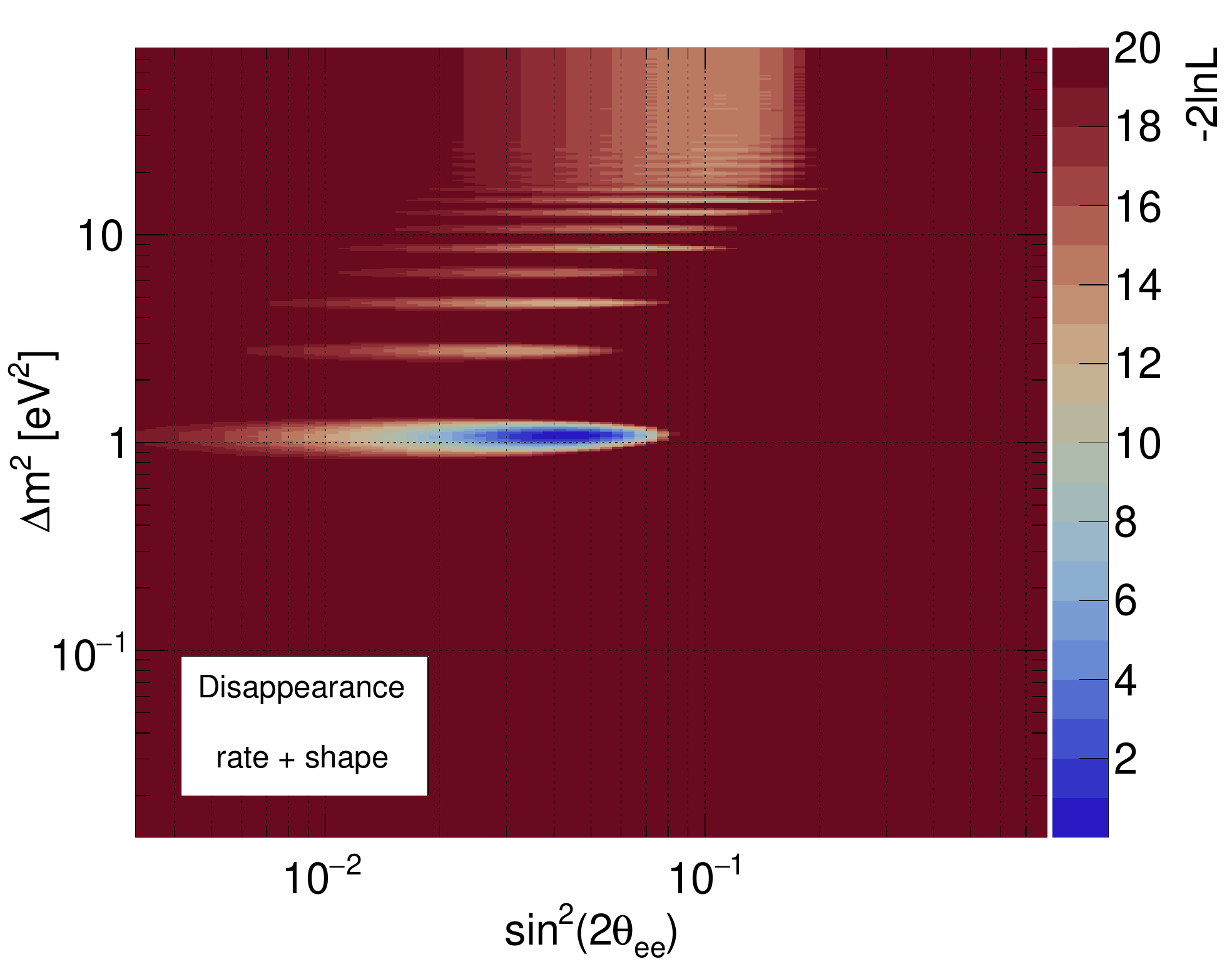}   
   \caption{Negative logarithm of the likelihood function for two sets of pseudo-data
      generated for the toy disappearance experiment, assuming the no-signal
      hypothesis (above) and the existence of a sterile neutrino
      with \sst=0.04 and \dms=1\,eV$^2$ (bottom). 
      Both maps are normalized to the absolute minimum in the \sst-\dms\ parameter space.}
   \label{fig:likelihood-map}
\end{figure}

\figurename ~\ref{fig:likelihood-map} shows the NLL
as a function of \dms\ and \sst\ for two sets of pseudo-data 
of the toy disappearance experiment. 
The two sets of pseudo-data are respectively a realization of the no-signal
hypothesis and a realization of a hypothesis with \sst=0.04 and \dms=1\,eV$^2$. 
They are the same data sets used for the comparison of the performance of the statistical
methods in the previous sections.
Local minima at regularly-spaced \dms\ values are present for both data sets. 
This feature is due to the oscillatory nature of the sought-after
signal and appears in any realizations of the data.

The presence of multiple minima makes it difficult for a minimization algorithm
to converge to the absolute minimum,
in particular for those algorithms relying on the derivative of the function (e.g.
the algorithms known as SIMPLEX and MIGRAD in the MINUIT software
package~\cite{James:1975dr}). 
To reliably find the absolute minimum we adopt a scanning
approach in which \dms\ is increased progressively with uniform steps in the
logarithmic space, each step having a length of $\log(\dms/\text{eV}^2)\lesssim
0.01$. 
At each \dms\ value a minimization against \sst\ is performed. This minimization
is not problematic because, when the value of \dms\ is fixed, the likelihood
function along \sst\ is a smooth function with a unique minimum.

The NLL for the pseudo-data generated under the
no-signal hypothesis shows another important feature: the absolute minimum does not
correspond to
the no-signal hypothesis. This is the case for all the realizations of the
data of the toy disappearance experiment. The sought-after oscillatory signature is indeed mimicked by the
statistical fluctuations between adjacent bins and the data are described always
better by an oscillation hypothesis than by the no-signal hypothesis. As
discussed later in \ref{sec:app-pdf}, this leads to a deformation of the
test statistic probability distribution.

 \section{Generation of Pseudo-Data}\label{sec:app-pseudo-data}

The generation of pseudo-data is performed with Monte Carlo techniques.  The
experimental parameters of the two toy experiments used in this work
are quoted in Section~\ref{sec:exp-intro} and \tablename~\ref{tab:exp}.
Pseudo-data for a specific  hypothesis $\text{H(X,Y)}:\{\sst, \dms: \sst=X,
\dms=Y\}$ are generated 
according to the probability distribution of signal ($\PDF^e_{ij}(\text{X,Y})$)
and background ($\PDF^b_{ij}$) events as a function of
the $i$-th bin in \lrec\ and $j$-th bin \erec.
To construct a set of pseudo-data, a Poisson random number is generated for each
$ij$ bin using as expectation 
\begin{eqnarray*}
   \lambda_{ij} = \tilde{N}^e(\text{X,Y})\cdot \PDF^e_{ij}(\text{X,Y})+\tilde{N}^b\cdot \PDF^b_{ij}
\end{eqnarray*}
where $\tilde{N}^e(\text{X,Y})$ is the expected number of neutrino interactions under the
hypothesis H(X,Y) and $\tilde{N}^b$ is the expected number of background events.
For each set of pseudo-data also the external constraints $\bar{N}^e(\text{X,Y})$ and
$\bar{N}^b$ (see equation~\eqref{eq:likelihood-tot}) are sampled from 
Gaussian distributions with means $\tilde{N}^e(\text{X,Y})$ and $\tilde{N}^b$ and standard deviations
$\sigma^e$ and $\sigma^b$ respectively.

$N^e$  and $N^b$ are nuisance parameters in our analysis and their true values $\tilde{N}^e(\text{X,Y})$ and $\tilde{N}^b$ are regarded as fixed during the construction of ensembles of pseudo-data. In contrast to our construction, the expectation values
of the nuisance parameters could also be sampled from a prior probability
distribution. 
Ensembles generated in this way can be used to construct the probability
distribution of a test statistic taking into account the systematic
uncertainties on the nuisance parameters~\cite{Tanabashi:2018oca}.
This construction leads to probability distributions that are the average over a
set of models, each model having different values of the nuisance parameters
and a weight proportional to the prior probability of those specific values.
This construction can be considered for \test{0}.
In contrast, the asymptotic probability distribution for test statistics based
on the profile likelihood ratio such as \test{1} and \test{2} does not depend on
the value of the nuisance parameters. 
To ease the comparison between test statistics, we 
generated the ensembles of pseudo-data always assuming fixed values of the
nuisance parameters.
 \section{Hypothesis Testing and Sensitivity} \label{sec:app-neyman}

The definition of the set of accepted/rejected hypotheses in this work is based on a
Neyman construction in which the ordering principle is defined by the test
statistic. The construction follows the steps outlined in this section.

Let \test{} be any of the three statistics in \tablename~\ref{tab:stat}.
\test{} is used to test a specific hypothesis $\text{H(X,Y)}:\{\sst, \dms: \sst=X,
\dms=Y\}$ given a set of data.
We will use the symbol $\test{X,Y}$ to make explicit which hypothesis is being
tested.  The set of hypotheses accepted or rejected by the test 
is defined according to how the observed value of the test statistic
$\test{X,Y}^\text{obs}$ compares to its 
critical value $\test{X,Y}^\text{crit}$.
To this purpose, the critical value is determined for a predefined test size $\alpha$ as:
\begin{eqnarray*}
\alpha = \int_{\test{X,Y}^\text{crit}}^{\infty} f(\test{X,Y}|\text{H(X,Y)}) \, d\test{X,Y}
\end{eqnarray*}
The probability distributions $f(\test{X,Y}|\text{H(X,Y)})$ are constructed by computing the test statistic
for each data set in ensembles of $10^4$ pseudo-data
sets produced under the true hypothesis H(X,Y).  

For each hypothesis to be tested the following steps are carried out:
\begin{enumerate}
   \item compute $\test{X,Y}^\text{crit}$ given the
      specific probability distribution for H(X,Y) and a test size of 0.05;
   \item $\test{X,Y}^\text{obs}$ is evaluated for the considered data set;
   \item the hypothesis is accepted if $\test{X,Y}^\text{obs} \leq \test{X,Y}^\text{crit}$, otherwise the hypothesis is rejected.
\end{enumerate}

Two definitions of sensitivity have been used in this article: the exclusion sensitivity
that defines the set of hypotheses that would be rejected
assuming the no-signal hypothesis, and the discovery sensitivity that
defines the set of hypotheses for which we expect to reject the no-signal
hypothesis.
The formal definition of these two quantities is based on the median
expected value of the test statistic under certain hypotheses that can be calculated using an ensemble of
pseudo-data.

The exclusion sensitivity is defined as the set of hypotheses
for which the test is expected to provide a median test statistic value exactly of $\test{X,Y}^\text{crit}$
assuming the data were generated from the no-signal hypothesis.
To find the set of hypotheses fulfilling this requirement, the
following steps are carried out for each H(X,Y):
\begin{enumerate}
   \item compute $\test{X,Y}^\text{crit}$ given the
      specific probability distribution for H(X,Y) and a test size of 0.05;
   \item compute $\text{med}[\test{X,Y}|\text{H(0,0)}]$, i.e. the median value of
      $f(\test{X,Y}|\text{H(0,0)})$;
\end{enumerate}
Since the set of hypotheses considered is usually discrete, 
finding a hypothesis that meets perfectly the condition above is not possible.
The requirement is
hence softened: for a given \dms\ value, the hypotheses are tested for
increasing values of \sst\ and the first one fulfilling the condition $\text{med}[\test{X,Y}|\text{H(0,0)}] > \test{X,Y}^\text{crit}$ is added to the set. 

The discovery sensitivity is defined as the set of hypotheses that, if
true, would result in a rejection of the no-signal hypothesis with a probability of 50\%.
To find the set of hypotheses fulfilling this requirement, the
following steps are carried out for each H(X,Y):
\begin{enumerate}
   \item $\test{0,0}^\text{crit}$ is computed given the
      specific probability distribution for H(0,0) and a test size of 0.05;
   \item compute $\text{med}[\test{0,0}|\text{H(X,Y)}]$, i.e. the median value of
      $f(\test{0,0}|\text{H(X,Y)})$;
\end{enumerate}
Similarly to the previous case, a softer requirement is typically applied when
considering a discrete set of hypotheses.
 \section{Probability Distributions of \test{2}}\label{sec:app-pdf}
\begin{figure*}[]
   \centering
   \includegraphics[width=1\textwidth]{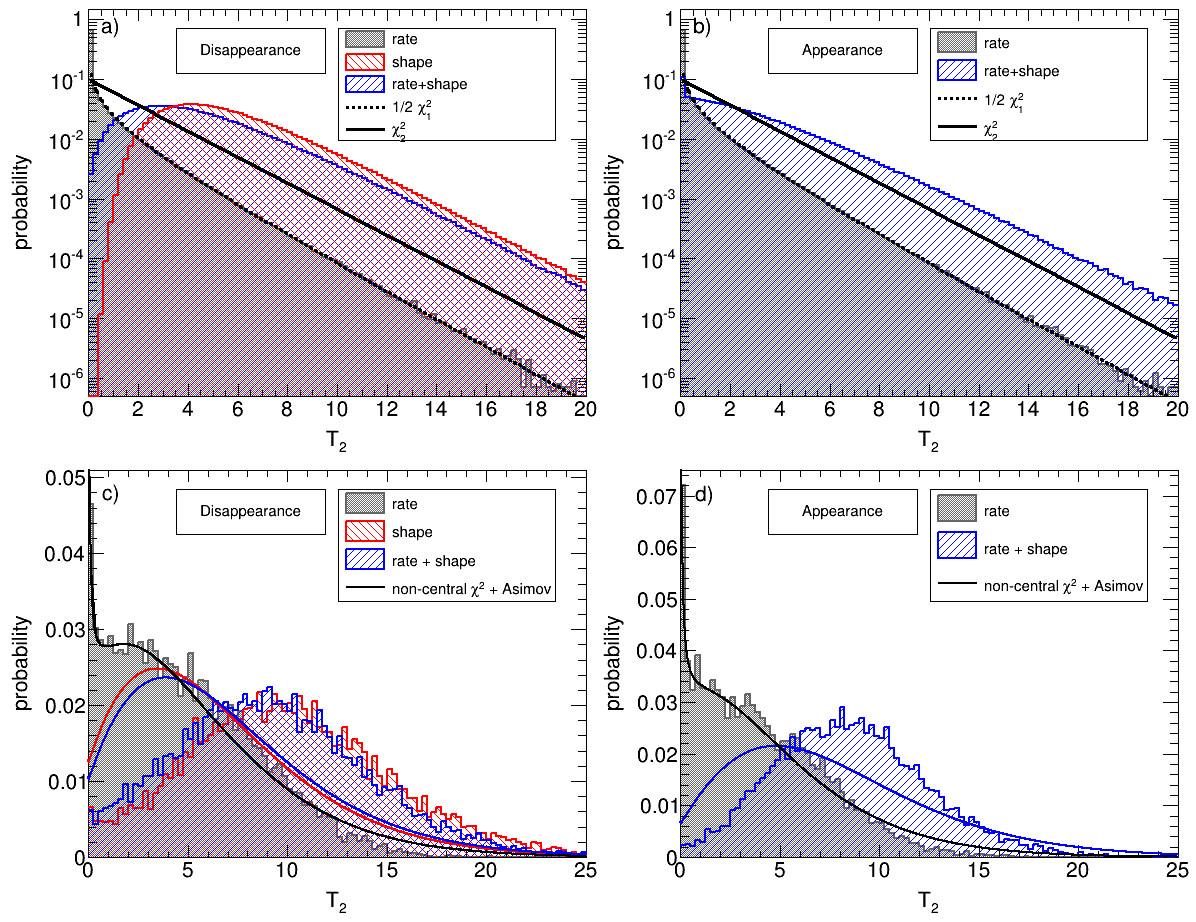}
   \caption{Distributions of \test{2} computed from an ensemble of pseudo-data
   generated assuming the no-signal hypothesis and their asymptotic formulas. 
   The distributions are shown for a test of the no-signal hypothesis (panel a
   and b using $10^7$ sets of pseudo-data) and of an oscillation hypothesis 
   (panel c and d using $10^4$ sets of pseudo-data). The oscillation
   hypothesis has \dms$=$1\,eV$^2$ and \sst$=$0.002/0.02/0.07, respectively for
   the appearance/disappearance(shape\&combined)/disappearance(rate) experiment. The value of the angle is chosen such that the
   hypothesis is very close to the experimental sensitivity for each kind of
   analysis.}
   \label{fig:teststat}
\end{figure*}

\figurename~\ref{fig:teststat} shows the probability distributions of \test{2}
computed from an ensemble of pseudo-data generated assuming the no-signal
hypothesis and their expected asymptotic formulas. 
The distributions are shown for the toy disappearance and appearance experiments,
assuming a rate analysis, a shape analysis, and their combination.

The top panels show the distributions obtained when testing the no-signal hypothesis
(i.e. testing the same hypothesis used to generate the pseudo-data).
In this case, according to Wilks' theorem, the distributions should tend
asymptotically to a chi-square function. However, considering that
\sst\ is bounded to positive values, the distributions are expected to be
described by a half chi-square~\cite{Chernoff:1954eli}.
The bottom panels show the distributions obtained when testing an oscillation
hypothesis (i.e. testing an hypothesis different from the one used to generate
the pseudo-data).
In this case, according to Wald~\cite{wald_43}, the distributions should
tend asymptotically to a non-central chi-square. 
In both cases the number of degrees of freedom of the chi-square function is
given by the difference between the number of free parameters
in the alternative and the null hypothesis. The non-centrality parameter is defined
through the Asimov data set as discussed in Ref.~\cite{Cowan:2010js}. 

The distributions for the rate analysis are to a first approximation described
by the asymptotic formulas.
On the contrary, the distributions for the shape analysis and the combination of
rate and shape differ significantly and are 
deformed towards higher values of the test statistic.
This is due to the fact that the sought-after oscillatory signature is often
preferred to the no-oscillation hypothesis and the best fit does not correspond
to the true hypothesis. 
In other words, the extra flexibility of a model with oscillations can always be
used to better describe the data and the statistical fluctuations between bins.

\begin{figure}[]
   \centering
   \includegraphics[width=\columnwidth]{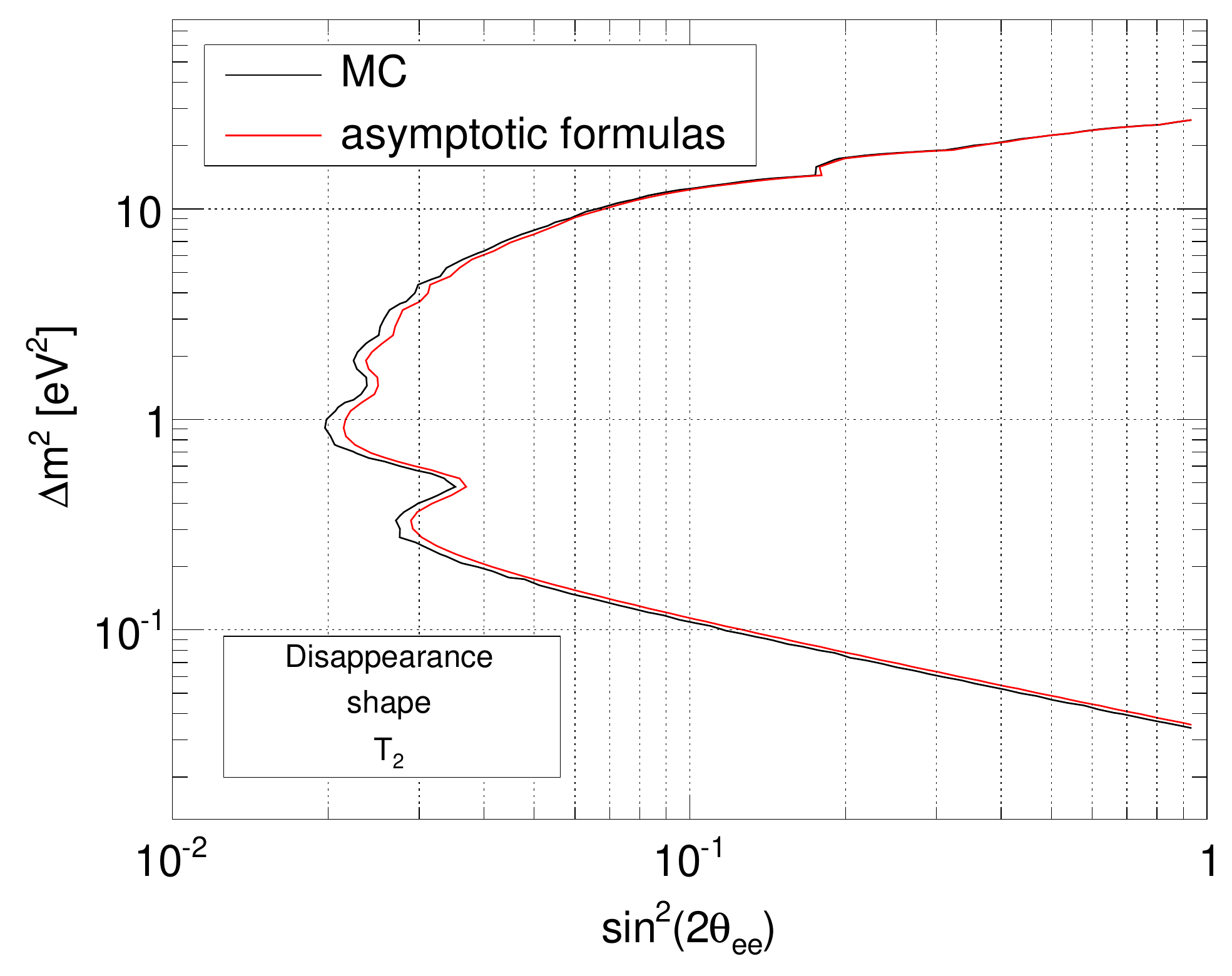}
   \includegraphics[width=\columnwidth]{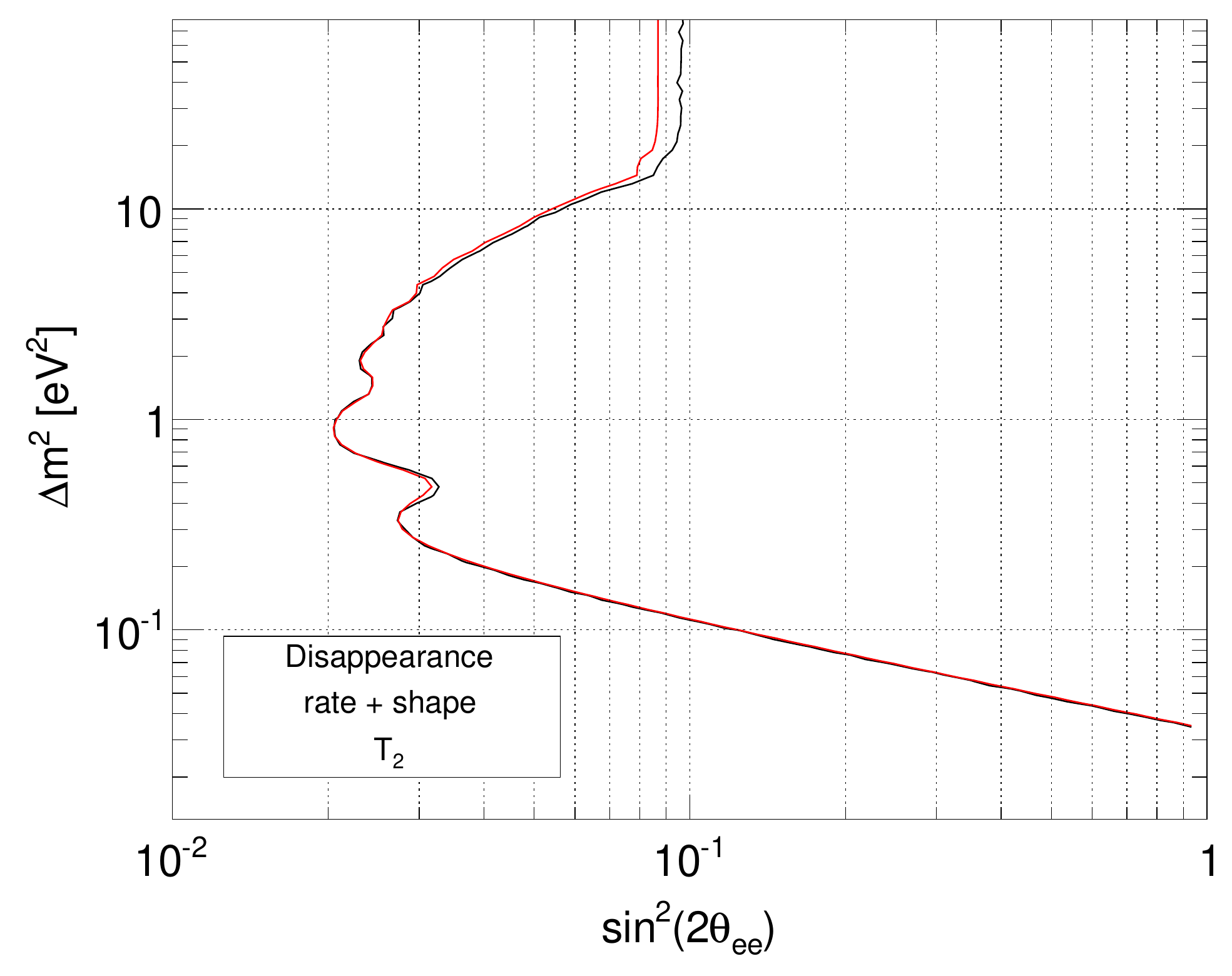}
   \includegraphics[width=\columnwidth]{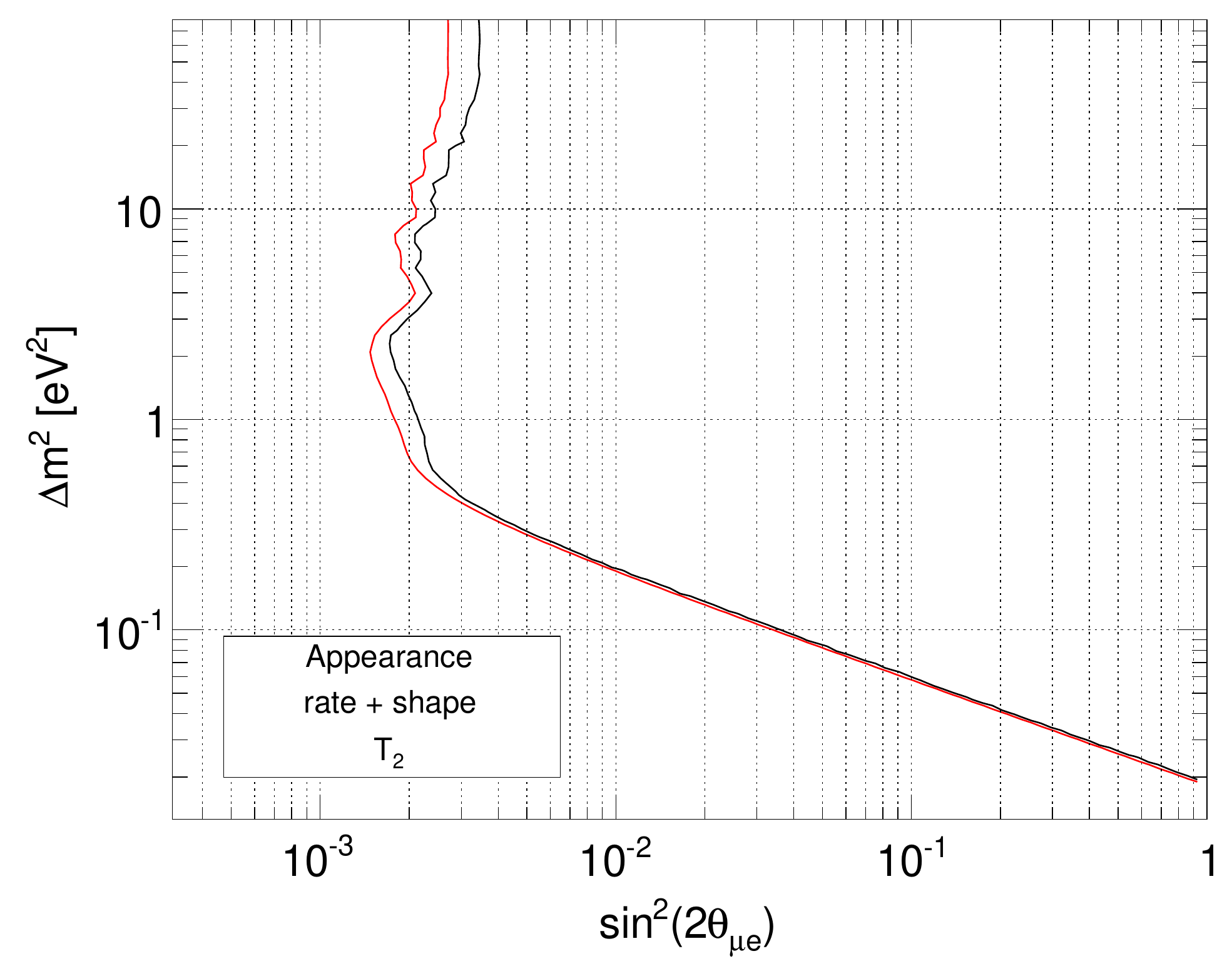}
   \caption{Comparison between the 95\% C.L. exclusion sensitivities computed
     using the test statistic distributions generated through Monte Carlo techniques
     and their analytical approximation based on Wilks and Wald's formulas and the Asimov
     data set.
     The sensitivities are shown for the toy disappearance and appearance
     experiment, using the shape and combined analysis. The sensitivities for a rate analysis fully agree with 
     each other and are not shown.}
   \label{fig:sensitivity-asimov}
\end{figure}

The exclusion sensitivities based on the asymptotic formulas are compared to
those based on the distributions constructed through Monte Carlo techniques in
\figurename ~\ref{fig:sensitivity-asimov}.
The approximated sensitivities calculated using the asymptotic formulas are
accurate within 10\%. This might seem inconsistent with the fact that 
the test statistic distributions are not well described by the asymptotic
formulas. However, the distributions are consistently shifted at higher values
for all possible hypotheses and this coherent bias preserves the relationship
between the distribution quantiles. 
This might not be valid if the sensitivity is
computed for an other confidence level (i.e. an other quantile of the distribution)
or if the experimental parameters are changed. 
 \section{Look-Elsewhere Effect}\label{sec:app-lee}
\begin{figure}[]
   \centering
   \includegraphics[width=\columnwidth]{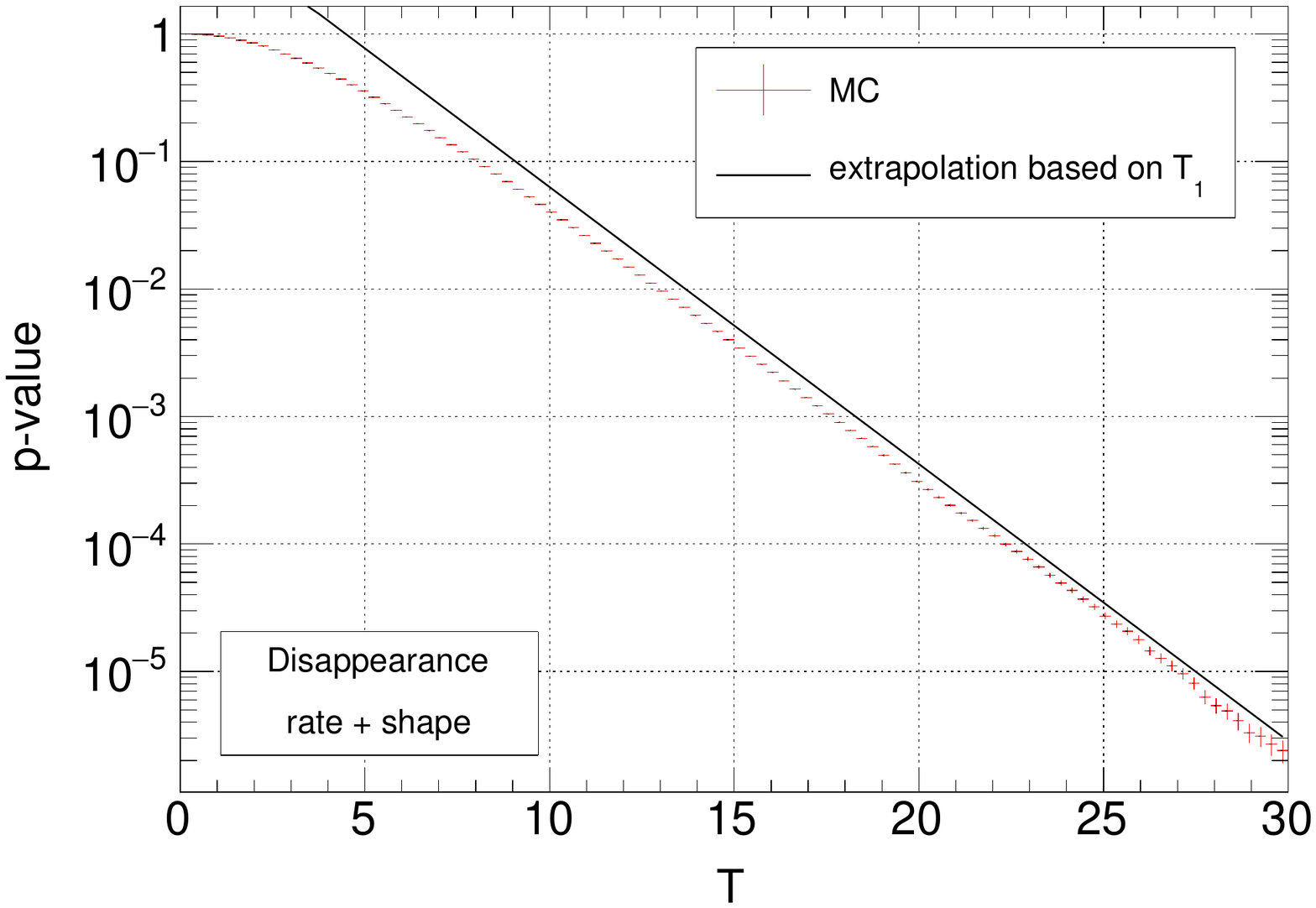}
   \caption{global p-value computed for the no-signal hypothesis as a function of the value of the
   test statistic \test{2}. The distribution computed from an ensemble of 
   pseudo-data generated under the no-signal hypothesis (MC) is compared with the
   extrapolation based on \test{1} and the procedure discussed in the text. The
   value of the parameters used for the extrapolation are $u_0 = 1$ and
   $\left<N_{u_0}\right> = 5.603$, as estimated from 1000 sets of pseudo-data.  }
   \label{fig:lee}
\end{figure}
The statistical significance associated to the observation of a sterile neutrino
signal can be expressed through the p-value computed for the no-signal hypothesis
given the observed data:
\begin{eqnarray*}
\pvalue{} = \int_{\test{0,0}^\text{obs}}^{\infty} f(\test{0,0}|\text{H(0,0)}) \, d\test{0,0}
\end{eqnarray*}
While \test{2} provides a unique p-value for the no-signal hypothesis, \test{1}
provides a p-value for each tested value of \dms\ (see
\figurename~\ref{fig:discovery}). 
For this reason the p-value provided by \test{1} is often called local
p-value, while the p-value provided by \test{2} is called global p-value.
A procedure to estimate the global p-value using the local estimation has been
proposed in Ref.~\cite{Gross:2010qma} based on previous results from
Davis~\cite{Davies:1987zz}. 
The procedure is based on a linear correction of the
minimum local p-value found:
\begin{equation*}
   \pvalue{global} \approx \min\limits_{\dms}\,\pvalue{local} + \left<N_u\right>
\end{equation*}
where $\left<N_u\right>$ is the mean number of ``upcrossings'' 
above the level $u$ in the range of considered \dms\ values for 
a test of the no-signal hypothesis based on \test{1}.
Each upcrossing corresponds to a \dms\ value for which the signal hypothesis
 is preferred over the no-signal hypothesis at a certain level $u$.
The mean number of upcrossings above the level $u$ and lower level $u_0$ is
connected by the relationship
$\left<N_u\right>= \left<N_{u_0}\right> e^{-(u-u_0)/2}$. 
$\left<N_{u_0}\right>$ can be estimated from a small
ensemble of pseudo-data. The possibility of using a small ensemble is
convenient because if $u_0$ is small  the number of upcrossings per
data set becomes large. 
The approximation becomes valid for $u\rightarrow\infty$, an upper limit on the p-value is given otherwise.  

Computing a global p-value from a local estimation could be 
interesting for sterile neutrino searches as the local p-value
construction can be performed assuming the asymptotic formula for the 
probability distributions of \test{1}.
The global p-value estimated through the correction discussed above is shown in \figurename~\ref{fig:lee} for the toy disappearance experiment. 
The p-value computed from \test{2} using an ensemble of pseudo-data is also shown.
The distributions have the
same shape for large values of the test statistic but a different offset. The
offset implies that the p-value extrapolated from \test{1} would be
overestimated by a factor of 1.4. This factor is not constant and depends on the
specific experiment.

While a correction can provide accurate
results in problems such as a peak search, the oscillatory nature of the signal
sought-after by sterile neutrino experiments induces a correlation in the number
of upcrossings. Such a correlation is due to the harmonics in
equation~\eqref{eq:osc-prob-dis} and equation~\eqref{eq:osc-prob-app} occurring
at different values of \dms.
It might be possible to correct the number of upcrossings to account for
the spurious occurrences but this requires additional studies.
 
\begin{acknowledgements}
The authors are greatly thankful to Hans Niederhausen for the valuable suggestions
on how to interpret the statistical methods and structure the comparison. 
The authors would also like to thank T.~Lasserre, P.~Litchfield, L.~Oberauer,
G.~Ranucci, M.~Wurm and S.~Schoenert for many helpful discussions and
suggestions during the work and the preparation of the manuscript.  This work
has been supported by the Deutsche Forschungsgemeinschaft (SFB1258).
\end{acknowledgements}

\providecommand{\href}[2]{#2}\begingroup\raggedright\endgroup
 

\begin{thebibliography}{10}

\bibitem{Giunti:2019aiy}
C.~Giunti and T.~Lasserre, \emph{{eV-scale Sterile Neutrinos}},
  \href{https://doi.org/10.1146/annurev-nucl-101918-023755}{\emph{Ann. Rev.
  Nucl. Part. Sci.} {\bfseries 69} (2019) 163--190}.

\bibitem{Pontecorvo:1967fh}
B.~Pontecorvo, \emph{{Neutrino Experiments and the Problem of Conservation of
  Leptonic Charge}}, {\emph{Sov. Phys. JETP} {\bfseries 26} (1968) 984--988}.

\bibitem{Boyarsky:2018tvu}
A.~Boyarsky, M.~Drewes, T.~Lasserre, S.~Mertens and O.~Ruchayskiy,
  \emph{{Sterile Neutrino Dark Matter}},
  \href{https://doi.org/10.1016/j.ppnp.2018.07.004}{\emph{Prog. Part. Nucl.
  Phys.} {\bfseries 104} (2019) 1--45}.

\bibitem{Aguilar:2001ty}
A.~Aguilar-Arevalo et~al. ({\scshape LSND} Collaboration), \emph{{Evidence for
  neutrino oscillations from the observation of $\bar{\nu}_e$ appearance in a
  $\bar{\nu}_\mu$ beam}},
  \href{https://doi.org/10.1103/PhysRevD.64.112007}{\emph{Phys. Rev. D}
  {\bfseries 64} (2001) 112007}.

\bibitem{Mention:2011rk}
G.~Mention, M.~Fechner, T.~Lasserre, T.~Mueller, D.~Lhuillier, M.~Cribier
  et~al., \emph{{The Reactor Antineutrino Anomaly}},
  \href{https://doi.org/10.1103/PhysRevD.83.073006}{\emph{Phys. Rev. D}
  {\bfseries 83} (2011) 073006}.

\bibitem{Giunti:2012tn}
C.~Giunti, M.~Laveder, Y.~Li, Q.~Liu and H.~Long, \emph{{Update of
  Short-Baseline Electron Neutrino and Antineutrino Disappearance}},
  \href{https://doi.org/10.1103/PhysRevD.86.113014}{\emph{Phys. Rev. D}
  {\bfseries 86} (2012) 113014}.

\bibitem{Dentler:2018sju}
M.~Dentler, A.~Hernandez-Cabezudo, J.~Kopp, P.~A. Machado, M.~Maltoni,
  I.~Martinez-Soler et~al., \emph{{Updated Global Analysis of Neutrino
  Oscillations in the Presence of eV-Scale Sterile Neutrinos}},
  \href{https://doi.org/10.1007/JHEP08(2018)010}{\emph{JHEP} {\bfseries 08}
  (2018) 010}.

\bibitem{Feldman:1997qc}
G.~J. Feldman and R.~D. Cousins, \emph{{A Unified approach to the classical
  statistical analysis of small signals}},
  \href{https://doi.org/10.1103/PhysRevD.57.3873}{\emph{Phys. Rev. D}
  {\bfseries 57} (1998) 3873--3889}.

\bibitem{Lyons:2014kta}
L.~Lyons, \emph{{Raster scan or 2-D approach?}},
  \href{https://arxiv.org/abs/arXiv:1404.7395}{{\ttfamily arXiv:1404.7395}}.

\bibitem{Qian:2014nha}
X.~Qian, A.~Tan, J.~J. Ling, Y.~Nakajima and C.~Zhang, \emph{{The Gaussian
  CL$_s$ method for searches of new physics}},
  \href{https://doi.org/10.1016/j.nima.2016.04.089}{\emph{Nucl. Instrum. Meth.
  A} {\bfseries 827} (2016) 63--78}.

\bibitem{Tanabashi:2018oca}
M.~Tanabashi et~al. ({\scshape Particle Data Group} Collaboration),
  \emph{{Review of Particle Physics}},
  \href{https://doi.org/10.1103/PhysRevD.98.030001}{\emph{Phys. Rev. D}
  {\bfseries 98} (2018) 030001}.

\bibitem{Alekseev:2018efk}
I.~Alekseev et~al. ({\scshape DANSS} Collaboration), \emph{{Search for sterile
  neutrinos at the DANSS experiment}},
  \href{https://doi.org/10.1016/j.physletb.2018.10.038}{\emph{Phys. Lett. B}
  {\bfseries 787} (2018) 56--63}.

\bibitem{Ko:2016owz}
Y.~Ko et~al. ({\scshape NEOS} Collaboration), \emph{{Sterile Neutrino Search at
  the NEOS Experiment}},
  \href{https://doi.org/10.1103/PhysRevLett.118.121802}{\emph{Phys. Rev. Lett.}
  {\bfseries 118} (2017) 121802}.

\bibitem{Serebrov:2018vdw}
A.~Serebrov et~al. ({\scshape NEUTRINO-4} Collaboration), \emph{{First
  Observation of the Oscillation Effect in the Neutrino-4 Experiment on the
  Search for the Sterile Neutrino}},
  \href{https://doi.org/10.1134/S0021364019040040}{\emph{Pisma Zh. Eksp. Teor.
  Fiz.} {\bfseries 109} (2019) 209--218}.

\bibitem{Ashenfelter:2018iov}
J.~Ashenfelter et~al. ({\scshape PROSPECT} Collaboration), \emph{{First search
  for short-baseline neutrino oscillations at HFIR with PROSPECT}},
  \href{https://doi.org/10.1103/PhysRevLett.121.251802}{\emph{Phys. Rev. Lett.}
  {\bfseries 121} (2018) 251802}.

\bibitem{Abreu:2020bzt}
Y.~Abreu et~al. ({\scshape SoLid} Collaboration), \emph{{SoLid: A short
  baseline reactor neutrino experiment}},
  \href{https://arxiv.org/abs/arXiv:2002.05914}{{\ttfamily arXiv:2002.05914}}.

\bibitem{Almazan:2018wln}
H.~Almazan et~al. ({\scshape STEREO} Collaboration), \emph{{Sterile Neutrino
  Constraints from the STEREO Experiment with 66 Days of Reactor-On Data}},
  \href{https://doi.org/10.1103/PhysRevLett.121.161801}{\emph{Phys. Rev. Lett.}
  {\bfseries 121} (2018) 161801}.

\bibitem{Barinov:2019vmp}
V.~Barinov, V.~Gavrin, V.~Gorbachev, D.~Gorbunov and T.~Ibragimova, \emph{{BEST
  potential in testing the eV-scale sterile neutrino explanation of reactor
  antineutrino anomalies}},
  \href{https://doi.org/10.1103/PhysRevD.99.111702}{\emph{Phys. Rev. D}
  {\bfseries 99} (2019) 111702}.

\bibitem{Borexino:2013xxa}
G.~Bellini et~al. ({\scshape Borexino} Collaboration), \emph{{SOX: Short
  distance neutrino Oscillations with BoreXino}},
  \href{https://doi.org/10.1007/JHEP08(2013)038}{\emph{JHEP} {\bfseries 08}
  (2013) 038}.

\bibitem{Gaffiot:2014aka}
J.~Gaffiot et~al., \emph{{Experimental Parameters for a Cerium 144 Based
  Intense Electron Antineutrino Generator Experiment at Very Short Baselines}},
  \href{https://doi.org/10.1103/PhysRevD.91.072005}{\emph{Phys. Rev. D}
  {\bfseries 91} (2015) 072005}.

\bibitem{Ajimura:2017fld}
S.~Ajimura et~al., \emph{{Technical Design Report (TDR): Searching for a
  Sterile Neutrino at J-PARC MLF (E56, JSNS2)}},
  \href{https://arxiv.org/abs/arXiv:1705.08629}{{\ttfamily arXiv:1705.08629}}.

\bibitem{Aguilar-Arevalo:2018gpe}
A.~Aguilar-Arevalo et~al. ({\scshape MiniBooNE} Collaboration),
  \emph{{Significant Excess of ElectronLike Events in the MiniBooNE
  Short-Baseline Neutrino Experiment}},
  \href{https://doi.org/10.1103/PhysRevLett.121.221801}{\emph{Phys. Rev. Lett.}
  {\bfseries 121} (2018) 221801}
  [\href{https://arxiv.org/abs/1805.12028}{{\ttfamily 1805.12028}}].

\bibitem{Antonello:2015lea}
M.~Antonello et~al. ({\scshape MicroBooNE, LAr1-ND, ICARUS-WA104}
  Collaboration), \emph{{A Proposal for a Three Detector Short-Baseline
  Neutrino Oscillation Program in the Fermilab Booster Neutrino Beam}},
  \href{https://arxiv.org/abs/arXiv:1503.01520}{{\ttfamily arXiv:1503.01520}}.

\bibitem{An:2016luf}
F.~P. An et~al. ({\scshape Daya Bay} Collaboration), \emph{{Improved Search for
  a Light Sterile Neutrino with the Full Configuration of the Daya Bay
  Experiment}},
  \href{https://doi.org/10.1103/PhysRevLett.117.151802}{\emph{Phys. Rev. Lett.}
  {\bfseries 117} (2016) 151802}.

\bibitem{Hellwig:2017xhq}
D.~Hellwig and T.~Matsubara ({\scshape Double Chooz} Collaboration),
  \emph{{Sterile Neutrino Search with the Double Chooz Experiment}},
  \href{https://doi.org/10.1088/1742-6596/888/1/012133}{\emph{J. Phys. Conf.
  Ser.} {\bfseries 888} (2017) 012133}.

\bibitem{Yeo:2017ied}
I.~S. Yeo ({\scshape RENO} Collaboration), \emph{{Search for sterile neutrinos
  at RENO}}, \href{https://doi.org/10.1088/1742-6596/888/1/012139}{\emph{J.
  Phys. Conf. Ser.} {\bfseries 888} (2017) 012139}.

\bibitem{Adamson:2017uda}
P.~Adamson et~al. ({\scshape MINOS+} Collaboration), \emph{{Search for sterile
  neutrinos in MINOS and MINOS+ using a two-detector fit}}, {\emph{Phys. Rev.
  Lett.} {\bfseries 122} (2019) 091803}.

\bibitem{Adamson:2017zcg}
P.~Adamson et~al. ({\scshape NOvA} Collaboration), \emph{{Search for
  active-sterile neutrino mixing using neutral-current interactions in NOvA}},
  \href{https://doi.org/10.1103/PhysRevD.96.072006}{\emph{Phys. Rev. D}
  {\bfseries 96} (2017) 072006}.

\bibitem{Aartsen:2017bap}
M.~Aartsen et~al. ({\scshape IceCube} Collaboration), \emph{{Search for sterile
  neutrino mixing using three years of IceCube DeepCore data}},
  \href{https://doi.org/10.1103/PhysRevD.95.112002}{\emph{Phys. Rev. D}
  {\bfseries 95} (2017) 112002}.

\bibitem{Aghanim:2018eyx}
N.~Aghanim et~al. ({\scshape Planck} Collaboration), \emph{{Planck 2018
  results. VI. Cosmological parameters}},
  \href{https://arxiv.org/abs/arXiv:1807.06209}{{\ttfamily arXiv:1807.06209}}.

\bibitem{Angrik:2005ep}
J.~Angrik et~al. ({\scshape KATRIN} Collaboration), \emph{{KATRIN design report
  2004}}, {\emph{FZKA-7090} (2005) }.

\bibitem{Gastaldo:2016kak}
L.~Gastaldo, C.~Giunti and E.~Zavanin, \emph{{Light sterile neutrino
  sensitivity of 163Ho experiments}},
  \href{https://doi.org/10.1007/JHEP06(2016)061}{\emph{JHEP} {\bfseries 06}
  (2016) 061}.

\bibitem{Beaujean:2011zza}
F.~Beaujean, A.~Caldwell, D.~Kollar and K.~Kroninger, \emph{{p-values for model
  evaluation}}, \href{https://doi.org/10.1103/PhysRevD.83.012004}{\emph{Phys.
  Rev. D} {\bfseries 83} (2011) 012004}.

\bibitem{Baker:1983tu}
S.~Baker and R.~D. Cousins, \emph{{Clarification of the Use of Chi Square and
  Likelihood Functions in Fits to Histograms}},
  \href{https://doi.org/10.1016/0167-5087(84)90016-4}{\emph{Nucl. Instrum.
  Meth.} {\bfseries 221} (1984) 437--442}.

\bibitem{casella2003statistical}
G.~Casella and R.~L. Berger, \emph{Statistical inference, Second Edition}.
  Thomson Learning, Australia Pacific Grove, CA, 2002.

\bibitem{ATLAS:2011tau}
{\scshape ATLAS, CMS, LHC Higgs Combination Group}, \emph{{Procedure for the
  LHC Higgs boson search combination in summer 2011}},
  {\emph{ATL-PHYS-PUB-2011-011, CMS-NOTE-2011-005} (2011) }.

\bibitem{Gross:2010qma}
E.~Gross and O.~Vitells, \emph{{Trial factors for the look elsewhere effect in
  high energy physics}},
  \href{https://doi.org/10.1140/epjc/s10052-010-1470-8}{\emph{Eur. Phys. J. C}
  {\bfseries 70} (2010) 525--530}.

\bibitem{Ranucci:2012ed}
G.~Ranucci, \emph{{The Profile likelihood ratio and the look elsewhere effect
  in high energy physics}},
  \href{https://doi.org/10.1016/j.nima.2011.09.047}{\emph{Nucl. Instrum. Meth.
  A} {\bfseries 661} (2012) 77--85}.

\bibitem{Wilks:1938dza}
S.~Wilks, \emph{{The Large-Sample Distribution of the Likelihood Ratio for
  Testing Composite Hypotheses}},
  \href{https://doi.org/10.1214/aoms/1177732360}{\emph{Annals Math. Statist.}
  {\bfseries 9} (1938) 60--62}.

\bibitem{stuart2009kendall}
A.~Stuart and K.~Ord, \emph{Kendall's Advanced Theory of Statistics: Volume 1:
  Distribution Theory}, Kendall's Advanced Theory of Statistics. Wiley, 2009.

\bibitem{Algeri:2019arh}
S.~Algeri, J.~Aalbers, K.~Dundas~Mora and J.~Conrad, \emph{{Searching for new
  physics with profile likelihoods: Wilks and beyond}},
  \href{https://arxiv.org/abs/arXiv:1911.10237}{{\ttfamily arXiv:1911.10237}}.

\bibitem{Cowan:2010js}
G.~Cowan, K.~Cranmer, E.~Gross and O.~Vitells, \emph{{Asymptotic formulae for
  likelihood-based tests of new physics}},
  \href{https://doi.org/10.1140/epjc/s10052-011-1554-0}{\emph{Eur. Phys. J. C}
  {\bfseries 71} (2011) 1554}.

\bibitem{Chernoff:1954eli}
H.~Chernoff, \emph{{On the Distribution of the Likelihood Ratio}},
  \href{https://doi.org/10.1214/aoms/1177728725}{\emph{Ann. Math. Stat.}
  {\bfseries 25} (1954) 573--578}.

\bibitem{Read:2000ru}
A.~L. Read, \emph{{Modified frequentist analysis of search results (The CL(s)
  method)}},  in \emph{{Workshop on confidence limits, CERN, Geneva,
  Switzerland, 17-18 Jan 2000: Proceedings}}, pp.~81--101, 8, 2000.

\bibitem{James:1975dr}
F.~James and M.~Roos, \emph{{Minuit: A System for Function Minimization and
  Analysis of the Parameter Errors and Correlations}},
  \href{https://doi.org/10.1016/0010-4655(75)90039-9}{\emph{Comput. Phys.
  Commun.} {\bfseries 10} (1975) 343--367}.

\bibitem{wald_43}
A.~Wald, \emph{Tests of statistical hypotheses concerning several parameters
  when the number of observations is large}, {\emph{Trans. Am. Math. Soc.}
  {\bfseries 54} (1943) 426-482}.

\bibitem{Davies:1987zz}
R.~B. Davies, \emph{{Hypothesis testing when a nuisance parameter is present
  only under the alternative}},
  \href{https://doi.org/10.1093/biomet/74.1.33}{\emph{Biometrika} {\bfseries
  74} (1987) 33--43}.

\end{thebibliography}
\end{document}